\documentclass{aa}

\usepackage{graphicx}   
\usepackage{txfonts}    

\usepackage{hyperref}
\hypersetup{colorlinks=true,linkcolor=blue,citecolor=blue,filecolor=blue,urlcolor=blue,}
\usepackage{supertabular,booktabs}
\usepackage{caption}
\newcommand{\orcid}[1]{\href{https://orcid.org/#1}{\includegraphics[width=10pt]{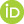}}}

\DeclareRobustCommand{\ion}[2]{%
\relax\ifmmode
\ifx\testbx\f@series
{\mathbf{#1\,\mathsc{#2}}}\else
{\mathrm{#1\,\mathsc{#2}}}\fi
\else\textup{#1\,{\mdseries\textsc{#2}}}%
\fi}
\usepackage[normalem]{ulem}

\begin{document}

        \title{Supernova environments in J-PLUS}
        \subtitle{Normalized cumulative-rank distributions and stellar-population synthesis combining narrow- and broad-band filters}

        \author{Ra\'ul Gonz\'alez-D\'iaz\inst{1,2}\thanks{Raul.GonzalezD@autonoma.cat}\orcid{0000-0002-0911-5141}
        \and
        Llu\'is Galbany\inst{1,3}\orcid{0000-0002-1296-6887}
        \and
        Tuomas Kangas\inst{4,5}\orcid{0000-0002-5477-0217}
        \and
        Rub\'en Garc\'ia-Benito\inst{6}\orcid{0000-0002-7077-308X}
        \and
        \mbox{Joseph~P. Anderson\inst{7,8}\orcid{0000-0003-0227-3451}}
        \and
        \mbox{Joseph Lyman\inst{9}}
        \and
        Jesús Varela\inst{10}
        \and
        Lamberto Oltra\inst{1}
        \and
        Rafael Logro\~no Garc\'ia\inst{10}
        \and
        \mbox{Gonzalo Vilella Rojo\inst{10}}
        \and
        \mbox{Carlos L\'opez-Sanjuan\inst{10}}
        \and
        Miguel \'Angel P\'erez-Torres\inst{6,15,16}
        \and
        Fabi\'an Rosales-Ortega\inst{2}
        \and
        \mbox{Seppo Mattila\inst{4,5}}
        \and
        \mbox{Hanindyo Kuncarayakti\inst{4,5}}
        \and
        Phil James\inst{11}
        \and
        Stacey Habergham\inst{11}
        \and
        Jos\'e Manuel V\'ilchez\inst{6}
\and \\ Jailson Alcaniz\inst{12}
\and \mbox{Raul E. Angulo\inst{13}}
\and Javier Cenarro\inst{10}
\and David Cristóbal-Hornillos\inst{10}
\and Renato Dupke\inst{12}
\and \mbox{Alessandro Ederoclite\inst{10}}
\and \mbox{Carlos Hernández-Monteagudo\inst{10}}
\and Antonio Marín-Franch\inst{10}
\and Mariano Moles\inst{10}
\and \mbox{Laerte Sodré Jr.\inst{14}}
\and Héctor Vázquez Ramió\inst{10}
}

   \institute{Institute of Space Sciences (ICE-CSIC), Campus UAB, Carrer de Can Magrans, s/n, E-08193 Barcelona, Spain. 
         \and
         Instituto Nacional de Astrof\'isica, \'Optica y Electr\'onica (INAOE-CONAHCyT), Luis E. Erro 1, 72840 Tonantzintla, M\'exico.
         \and
            Institut d’Estudis Espacials de Catalunya (IEEC), E-08034 Barcelona, Spain.
         \and
             Finnish Centre for Astronomy with ESO (FINCA), FI-20014 University of Turku, Finland
          \and   
             Tuorla Observatory, Department of Physics and Astronomy, FI-20014 University of Turku, Finland
         \and
             Instituto de Astrof\'isica de Andaluc\'ia (CSIC), Glorieta de la Astronom\'ia s/n, Aptdo. 3004, E-18080-Granada, Spain.
         \and
             European Southern Observatory, Alonso de C\'ordova 3107, Casilla 19, Santiago, Chile.
         \and
          Millennium Institute of Astrophysics MAS, Nuncio Monsenor Sotero Sanz 100, Off. 104, Providencia, Santiago, Chile.
         \and
          Department of Physics, University of Warwick, Coventry CV4 7AL, UK.
         \and
          Centro de Estudios de F\'isica del Cosmos de Arag\'on (CEFCA), Unidad Asociada al CSIC, P. San Juan, 1, 44001 Teruel, Spain
          \and
          Astrophysics Research Institute, Liverpool John Moores University, IC2, 146 Brownlow Hill, Liverpool L3 5RF, UK
\and Observatorio Nacional, Rua Gal. Jos\'e Cristino 77, Rio de Janeiro, 20921-400, RJ, Brazil.
\and Donostia International Physics Center, Paseo Manuel de Lardizabal 4, 20018, Donostia-San Sebastian, Spain.
\and Instituto de Astronomia, Geof\'isica e Ciencias Atmosf\^ericas, U. S\~ao Paulo, R. do Mat\~ao 1226, S\~ao Paulo, 05508-090, SP, Brazil. 
\and Center for Astroparticles and High Energy Physics (CAPA), Universidad de Zaragoza, E-50009 Zaragoza, Spain.
\and School of Sciences, European University Cyprus, Diogenes street, Engomi, 1516 Nicosia, Cyprus.
        }

   \date{Received ... ; accepted ... }

\abstract{
We investigated the local environmental properties of 418 supernovae (SNe) of all types using data from the Javalambre Photometric Local Universe Survey (J-PLUS), which includes five broad-band and seven narrow-band imaging filters. Our study involves two independent analyses: 1) the normalized cumulative-rank (NCR) method, which utilizes all 12 single bands along with five continuum-subtracted narrow-band emission and absorption bands, and 2) simple stellar population (SSP) synthesis, where we build spectral energy distributions (SED) of the surrounding 1 kpc$^2$ SN environment using the 12 broad- and narrow-band filters.
Improvements on previous works include: (i) the extension of the NCR technique to other filters (broad and narrow) and the use a set of homogeneous data (same telescope and instruments); (ii) a correction for extinction to all bands based on the relation between the $g-i$ color and the color excess $E(B-V)$; and (iii) a correction for the contamination of the [\ion{N}{ii}] $\lambda$6583 line that falls within the H$\alpha$ filter.
All NCR distributions in the broad-band filters, tracing the overall light distribution in each galaxy, are similar to each other. The main difference is that type Ia, II, and IIb SNe are preferably located in redder environments than the other SN types. The radial distribution of the SNe shows that type IIb SNe seem to have a preference for occurring in the inner regions of galaxies, whereas other types of SNe occur throughout the galaxies without a distinct preference for a specific location.
For the H$\alpha$ filter we recover the sequence from SNe Ic, which has the highest NCR, to SNe Ia, which has the lowest; this is interpreted as a sequence in progenitor mass and age.
All core-collapse SN types are strongly correlated to the [\ion{O}{II}] emission, which also traces star formation rate (SFR), following the same sequence as in H$\alpha$.
The NCR distributions of the Ca II triplet show a clear division between II-IIb-Ia and Ib-Ic-IIn subtypes, which is interpreted as a difference in the environmental metallicity.
Regarding the SSP synthesis, we found that including the seven J-PLUS narrow filters in the fitting process has a more significant effect on the core-collapse SN environmental parameters than for SNe Ia, shifting their values toward more extincted, younger, and more star-forming environments, due to the presence of strong emission lines and stellar absorptions in those narrow bands.
}


\maketitle


\section{Introduction}

Supernovae (SNe) are one of the final stages in stellar evolution and are key in driving the chemical evolution of galaxies. 
Classical SNe (excluding superluminous ones) are basically divided in two main types: thermonuclear (those of type Ia) and core-collapse (CC).
Type Ia SNe are those triggered by the thermonuclear explosion of a carbon and oxygen white dwarf (WD;  \citealt{1960ApJ...132..565H}). Different progenitor scenarios, such as WDs reaching the Chandrasekhar limit ($M_{Ch}\sim 1.44M_{\odot}$) by mass accretion in a binary system or mergers of WDs, and explosion mechanisms, such as internal detonations of $M_{Ch}$, WD, or surface explosions on WDs with masses lower than $M_{Ch}$, have been proposed to explain how the WDs explode (see reviews by, e.g., \citealt{2014ARA&A..52..107M}). 
No direct progenitor detection has been reported to date (however, see \citealt{2014Natur.512...54M}), but in all cases, the main spectral features are the lack of H lines and the presence of strong Si II lines up to a couple of weeks after maximum brightness \citep{1997ARA&A..35..309F}. 
SNe Ia occur in galaxies of all types, including elliptical galaxies which only contain old stellar populations, providing strong evidence that SNe Ia have long-lived, low-mass progenitors \citep{2010ApJ...724..502H}. 
SNe Ia are more luminous than CC SNe, comprising about 30\% of the observed SNe \citep{2017ApJ...837..121G}. Their distinctive feature lies in the fact that generally normal SNe Ia exhibit similar spectra and light curves, making them valuable for tracing cosmological distances (e.g., \citealt{1999AJ....118.1766P, 1996ApJ...473...88R, 1997ApJ...483..565P, 1999ApJ...517..565P, 2006A&A...447...31A}).

Core collapse SNe are those resulting from the gravitational collapse of the iron core of a massive star ($>8 M_\odot$; \citealt{1989ApJ...339L..25A}). 
CC SNe are divided into three main subtypes depending on their spectral features, which reflect the state of the outer layers of the progenitor star at the moment of explosion. 
Type II SNe show H lines because their progenitors have kept the H-rich outer envelope intact, type Ib SNe show He but no H since the progenitor has lost the H envelope, and type Ic SNe lack both H and He lines due to the loss of both H- and He-rich layers \citep{1997ARA&A..35..309F}. 
Additionally, SNe IIn show narrow H emission lines most probably due to interaction with circumstellar material \citep{2011MNRAS.412.1522S}, and SNe IIb are those intermediate between SNe II and Ib, that show H only for a few days after explosion indicating that there was still a thin H layer before explosion \citep{1988AJ.....96.1941F, 1993Natur.364..507N}. 
There is a debate on the process responsible for the mass-loss leading to stripped envelope (SE) SN types (Ib, Ic, IIb). Mass-loss through radiation-driven wind from single hot massive stars or the removal of the outer envelope via tidal stripping by a less massive companion in a binary are the two most viable possibilities \citep{1997ARA&A..35..309F, 2017hsn..book..195G, 2017MNRAS.469.2672P, 2017PASP..129e4201S, 2018A&A...609A.136T}.

One outstanding question in the field of massive stellar evolution concerns the link between the nature of the progenitor and the resultant SN. Direct constraints on the nature of SN progenitors are limited. The bulk of these come from the small number of nearby events where deep, high-resolution (usually from the Hubble Space Telescope) pre-explosion imaging exists to directly image the progenitor star (e.g., \citealt{2013MNRAS.431L.102M,2015PASA...32...16S,2017RSPTA.37560277V}). For almost all discovered SNe, this is unfeasible. As such, methods investigating statistical properties of SNe have been developed to exploit the large numbers of observed SN environments (pre-SN or once the SN has faded), including historical ones. One of those statistical methods is the normalised cumulative-rank (NCR; \citealt{2006Natur.441..463F,2006A&A...453...57J}) method. 
The NCR method, when applied to H$\alpha$ narrow imaging, measures the correlation between SN locations and star-formation intensity (i.e., H$\alpha$ emission) in their host galaxies. The strength of this correlation is an indicator of the progenitor lifetime, a proxy for the initial mass of the progenitor.
The method has been employed on samples of SNe to deduce a sequence of ascending mass for the common SN types Ia, II, Ib, and Ic \citep{halphaem,2012MNRAS.424.1372A,2013MNRAS.436.3464K}. Subsequently, \citet{2017A&A...597A..92K} established a connection between the NCRs of these SNe and those of massive stars in the LMC and M33. This link indicated initial masses of approximately $>20 M_\odot$ for Ic SNe progenitors and $>9 M_\odot$ for II and Ib SNe.
For the stripped-envelope SN types, Ic and Ib, these findings would imply that single Wolf-Rayet stars and/or massive binaries are the most probable progenitors for the former, while for the latter interacting relatively low-mass binary systems, such as the directly detected 10$-$12 M$_\odot$ progenitor of the SN Ib iPTF13bvn \citep{2016MNRAS.461L.117E}, would be dominant.

Another method that has proved useful in putting constraints on SN progenitors is the study of the stellar populations at SN locations. \cite{2014A&A...572A..38G,2016A&A...591A..48G} presented the first statistical study of nearby SN host galaxies using integral field spectroscopy. These were provided by the CALIFA survey, which consists of 132 SNe of all types in 115 galaxies, and it was later extended to 272 SNe in 232 galaxies in  \cite{2018ApJ...855..107G}. Stellar parameters were inferred by fitting a set of single stellar population (SSP) models to the spectra of the locations where SNe occurred. They found that CC SNe tend to explode at positions with younger stellar populations than the galaxy average, while at Ia SN locations local properties were, on average, the same as global ones. They also found a sequence from higher to lower metallicity, from SN Ia to SN Ic-BL, and a significant increasing relative number of SNe Ic at higher metallicities compared to other CC SNe types, which supports a large fraction of this SN subtype occurring under the single-star scenario  (see also \citealt{2013AJ....146...30K,2018A&A...613A..35K}).

\begin{table}
\centering
\caption{J-PLUS photometric system properties. Transmission curves can be found at the J-PLUS website: \href{http://www.j-plus.es/survey/instrumentation}{http://www.j-plus.es/survey/instrumentation}.}
\label{tjplus}
\resizebox{\columnwidth}{!}{%
\begin{tabular}{llll}
\hline
Filter name & Central $\lambda$ (nm) & FWHM (nm) & Comments \\ \hline
uJAVA & 348.5 & 50.8 & ultraviolet continuum \\
J0378 & 378.5 & 16.8 & [OII] emission\\
J0395 & 395.0 & 10.0 & Ca H+K absorption \\
J0410 & 410.0 & 20.0 & H$\delta$ emission\\
J0430 & 430.0 & 20.0 & G-band absorption\\
gSDSS & 480.3 & 140.9 & SDSS green continuum \\
J0515 & 515.0 & 20.0 & Mgb Triplet absorption\\
rSDSS & 625.4 & 138.8 & SDSS red continuum \\
J0660 & 660.0 & 14.5 & H$\alpha$ emission\\
iSDSS & 766.8 & 153.5 & SDSS near infrared continuum\\
J0861 & 861.0 & 40.0 & Ca II triplet absorption\\
zSDSS & 911.4 & 140.9 & SDSS near infrared continuum\\ \hline
\end{tabular}%
}
\end{table}

\begin{figure}
    \centering
    \includegraphics[width=\columnwidth]{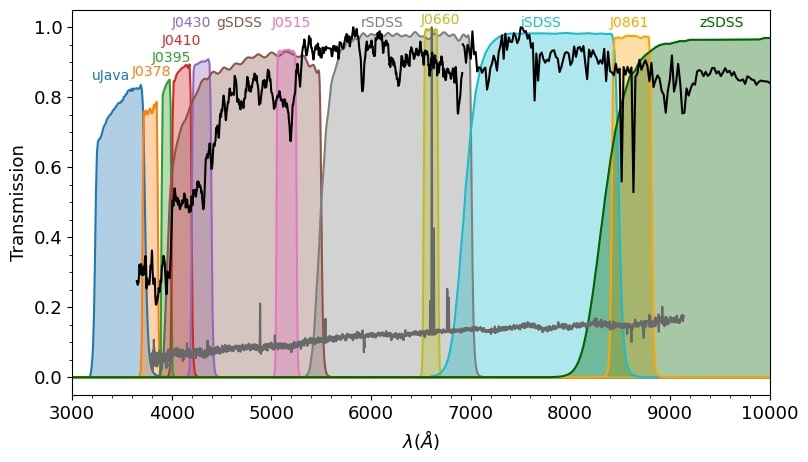}
    \caption{Transmission of 12 J-PLUS filters, on top of a typical star-forming (in gray) and passive galaxy (in black) spectra. Labels, which represent the filter names, highlight the main spectral features covered by the intermediate and narrow band filters.}
    \label{fig:filters}
\end{figure}

\begin{table*}
\centering
\caption{Pivot wavelengths and zero points of the J-PLUS bands.}
\resizebox{\textwidth}{!}{
\begin{tabular}{lcccccc}
\hline
Filter                     & uJAVA  & J0378  & J0395  & J0410  & J0430  & gSDSS  \\ \hline
$\lambda_{\rm pivot}$ (nm) & 352.29 & 378.64 & 395.06 & 410.07 & 430.04 & 474.46 \\
ZP (AB mag)                & 21.078 (049) & 20.416 (036) & 20.332 (037) & 21.283 (029) & 21.380 (028) & 23.553 (035)\\ \hline
Filter                     & J0515  & rSDSS  & J0660  & iSDSS  & J0861  & zSDSS  \\ \hline
$\lambda_{\rm pivot}$ (nm) & 514.98 & 622.98 & 659.98 & 767.65 & 860.25 & 892.20 \\
ZP (AB mag)                & 21.500 (040) & 23.494 (023) & 20.953 (023) & 23.136 (023) & 21.439 (007) & 22.498 (007)\\ \hline
\end{tabular}}
\label{Tzp}
\end{table*}

In this paper, we present a comprehensive analysis of 418 SNe environments using the 12 filter system of the Javalambre Photometric Local Universe Survey (J-PLUS\footnote{\url{www.j-plus.es}}; \citealt{2019A&A...622A.176C}), whose spatial resolution ($\lesssim$ 382 pc, corresponding to an angular resolution of 1.14 arcsec at z = 0.016)  and overall data quality of this dataset allow us to study the SN environments. Our investigation involves two main methodologies: the normalized cumulative-rank (NCR) analysis and the simple stellar population (SSP) synthesis. Compared to previous works in the literature, our study offers several advancements. Firstly, we extend the NCR technique to a number of broad- and narrow-band filters other than $H\alpha$, which enables us to explore correlations with new spectral lines. This extension opens up possibilities to probe additional environmental factors, such as stellar population metallicity or age. Previous studies, such as \citet[with UV imaging,]{2013MNRAS.436.3464K} and \citet[with NIR imaging]{2015PASA...32...19A}, have explored these factors to some extent. However, our research operates on a much larger scale and utilizes a set of homogeneous data acquired from the same telescope and instrument.
Another improvement over past works is the correction for the contamination from the [\ion{N}{ii}] $\lambda$6583 line that falls within the H$\alpha$ filter. \cite{dustandNII} presented a method to correct for such extra emission by combining narrow- and broad-band data and providing pure H$\alpha$ emission from J-PLUS observations. This method can also be extended to other narrow-band filters.


\section{Data}

\subsection{J-PLUS}

The Javalambre Photometric Local Universe Survey is designed to observe $8500$ deg$^2$ of the northern sky from the Observatorio Astrof\'{\i}sico de Javalambre (OAJ, Teruel, Spain; \citealt{oaj}) with the 83\,cm Javalambre Auxiliary Survey Telescope (JAST80) and T80Cam, a panoramic camera of 9.2k $\times$ 9.2k pixels that provides a $2\deg^2$ field of view (FoV) with a pixel scale of 0.55 arsec pix$^{-1}$ \citep{t80cam}. The J-PLUS filter system comprises the 12 broad, intermediate and narrow-band optical filters. 
J-PLUS is particularly designed to carry out the photometric calibration of the Javalambre Physics of the Accelerating Universe Astrophysical Survey (J-PAS\footnote{\url{http://www.j-pas.org/survey}}; \citealt{2014arXiv1403.5237B}),  which has observed the northern sky from Javalambre using 59 optical narrow-band filters. For this reason, some J-PLUS filters are located at key stellar spectral features that allow us to retrieve very accurate spectral energy distributions (SED) for more than five millions of stars in our galaxy.
The 12 filters of J-PLUS are listed in Table \ref{tjplus} and their transmission is shown in Figure \ref{fig:filters}, together with two template spectra of a star-forming and a passive galaxies to highlight the regions of interest covered by the intermediate- and narrow-band filters.
The third J-PLUS Data Release (DR3\footnote{\url{http://www.j-plus.es/datareleases/data_release_dr3}}) was made public in December 2022, comprising 1642 pointings observed in the 12 optical bands amounting to $\sim$3200 sq. deg with around 30 million sources detected at magnitudes $r<21$. 
We cross-matched all SN coordinates from the Open Supernova Catalogue (OSC; \citealt{2017ApJ...835...64G}) to the central coordinates of all J-PLUS DR3 tiles, and found 2168 SNe positions within the 2 sq. deg of the J-PLUS tile.
The width of the narrow H$\alpha$ filter (J0660 filter; $\sim$150 \AA) puts an upper limit on the redshift of galaxies that are useful for an NCR study at z$\sim$0.0163 (or about 60 Mpc), since the J0660 transmission falls down dramatically at 6672\AA. 
Thus, in order to keep the H$\alpha$ emission of the SN host galaxy within the coverage of the J-PLUS J0660 filter, we excluded all objects with a redshift larger than 0.0163. Moreover, NCR becomes less useful with distance, and samples at significantly different median distances may not be directly comparable (\citealt{2017A&A...597A..92K}). This redshift cut is passed by 282 SNe of the following types: 88 SNe Ia, 126 SNe II, 7 SNe IIn, 22 SNe Ib, 17 SNe Ic, 17 SNe IIb, and another five type Ibc SNe. 




\subsection{A dedicated SN host galaxy program at JAST80}

To reach a significantly large number of SNe that cover all types compared to previous studies, and within the redshift limit imposed by the J-PLUS H$\alpha$ filter width, we had to include objects outside the J-PLUS footprint, taking special care to increase the number of less represented subtypes (Ic, Ib, IIb, IIn). 
With this objective in mind, we initiated a dedicated program to acquire 73 additional fields using the same instrumental configuration and exposure times as in J-PLUS, which involves T80 and 12 filters. These fields encompassed the positions and host galaxies of 136 SNe. For each of the 12 filters, we obtained three individual frames, all of which were subsequently processed using the same J-PLUS pipeline \citep{2019A&A...622A.176C}.

Summing up the observations from the main survey and our dedicated program, 418 SNe are included in our sample, and they are listed in Table \ref{tab:SNetable}.
Taking into account the main SN types, we have 156 SNe II, 20 SN IIn, 108 SN Ia, 51 SN Ib, 49 SN Ic, 27 SN IIb, and seven other type Ibc SNe. The analysis is carried out with the sample divided into these seven main SN types, where Ibc SNe include Ib, Ic, IIb, and other Ibcs.               


\section{Generation of individual 3D data cubes}

Starting from the initial 12 images per field, one for each filter, we performed a number of steps to produce one 3D data cube per galaxy, namely galaxy cutout, flux calibration, correction for dust extinction, and continuum subtraction. All these steps are described in the following subsections. 

\subsection{Galaxy cutouts}

\begin{figure}
\centering
\includegraphics[width=\columnwidth]{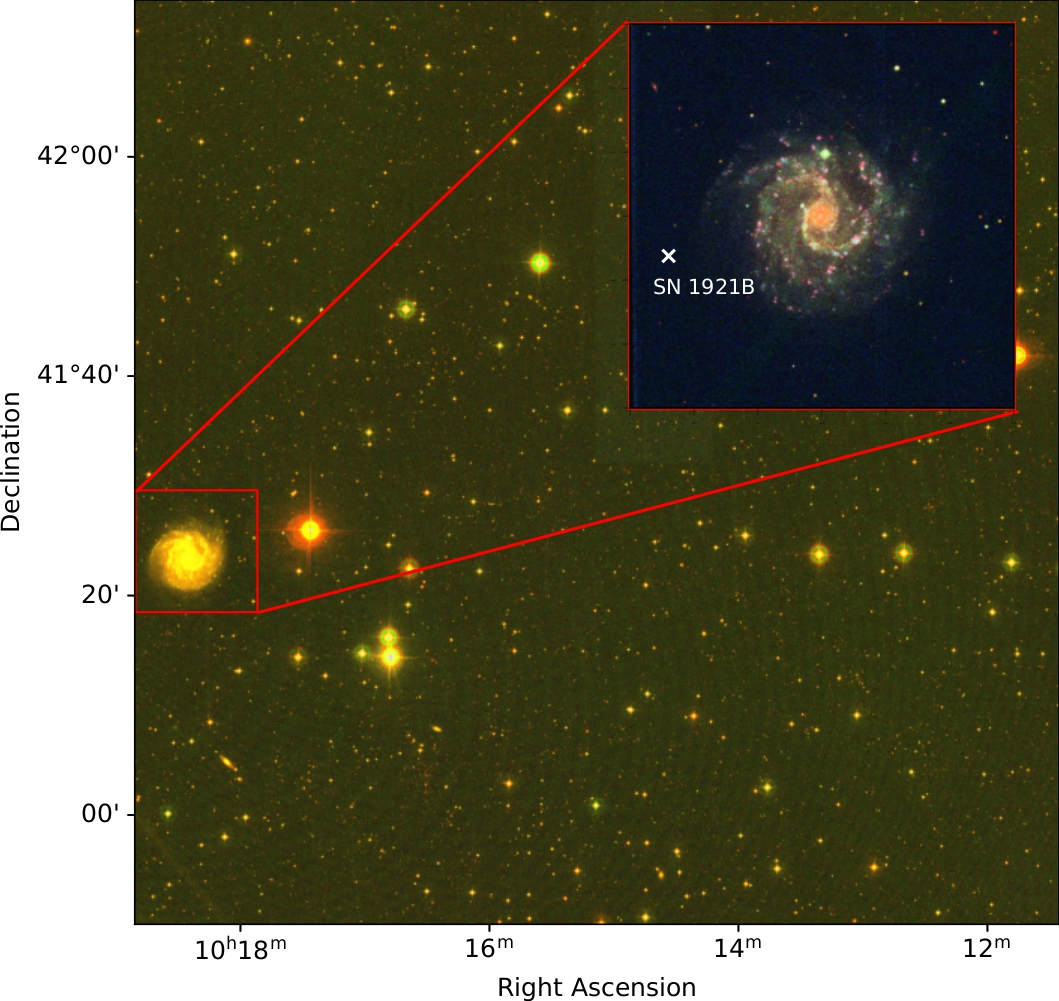}
\caption{Example of J-PLUS field frame. The zoomed-in image corresponds to the cutout of the galaxy NGC 3184, which contains the SN 1921B; marked in white. The images were constructed combining the uJAVA (blue), gSDSS (green), and rSDSS (red) images as false colors.}
\label{frame}
\end{figure}

J-PLUS images cover $\sim$2 sq. deg of the sky with a pixel scale of 0.55 arcsec, with a median seeing of 1.14 arcsec, corresponding to $\lesssim$ 382 pc at z = 0.016. After registering all 12 images to the same reference frame, we took a squared cutout of a size that includes the full extent of the SN host galaxy centered at its core, defining the size by visual inspection (see Fig. \ref{frame}).
The 12 cutouts are then stored in a 3D cube, mimicking integral field spectroscopy observations, where two spatial dimensions correspond to RA and Dec, and the third dimension provides a spectral energy distribution (SED) at any position within the cutout.

\subsection{Flux calibration}

To convert the electronic counts stored in each pixel to radiative flux, we use the expression from \cite{haJPLUS}:
\begin{equation}
    \centering
    F_\lambda=C\cdot10^{-0.4(ZP+48.6)}\frac{c}{\lambda^2_{\rm pivot}},
    \label{fluxec}
\end{equation}
where $C$ is the number count in the pixel, $c$ is the speed of light, $ZP$ is the zero point of the band used for the calibration to the standard AB magnitude system, and $\lambda_{\rm pivot}$ is the pivot wavelength of the filter, which is a source-independent measurement of the characteristic wavelength of a given pass band by
\begin{equation}
    \lambda^2_{pivot}=\frac{\int T(\lambda)d\lambda}{\int T(\lambda)\lambda^{-2} d\lambda},
\end{equation}
where $T(\lambda)$ represents the transmission curve of the filter.
%
The values of ZP and pivot wavelength are given in Table \ref{Tzp}.

\begin{figure*}
\centering
\includegraphics[width=0.4\textwidth]{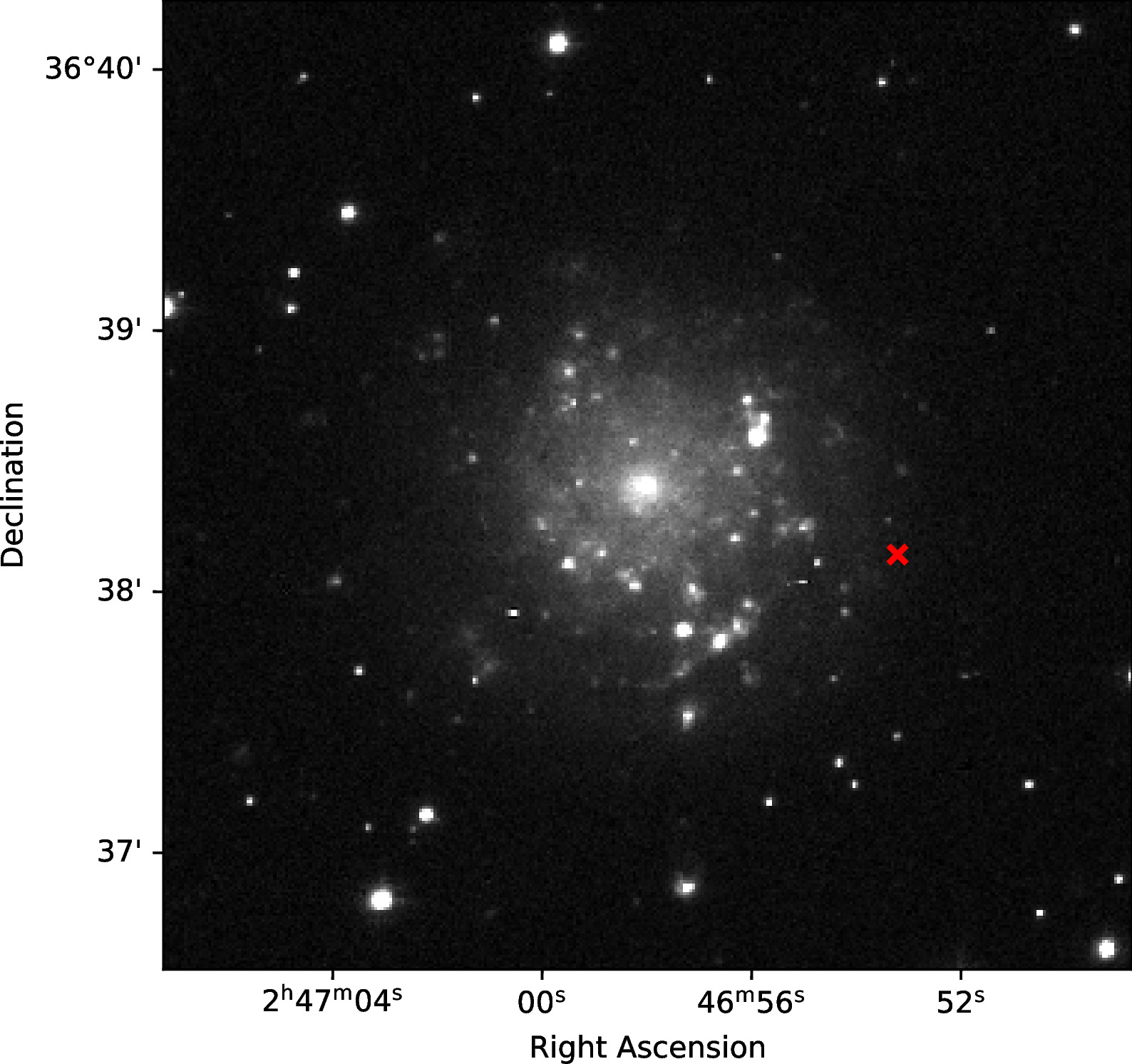}
\includegraphics[width=0.4\textwidth]{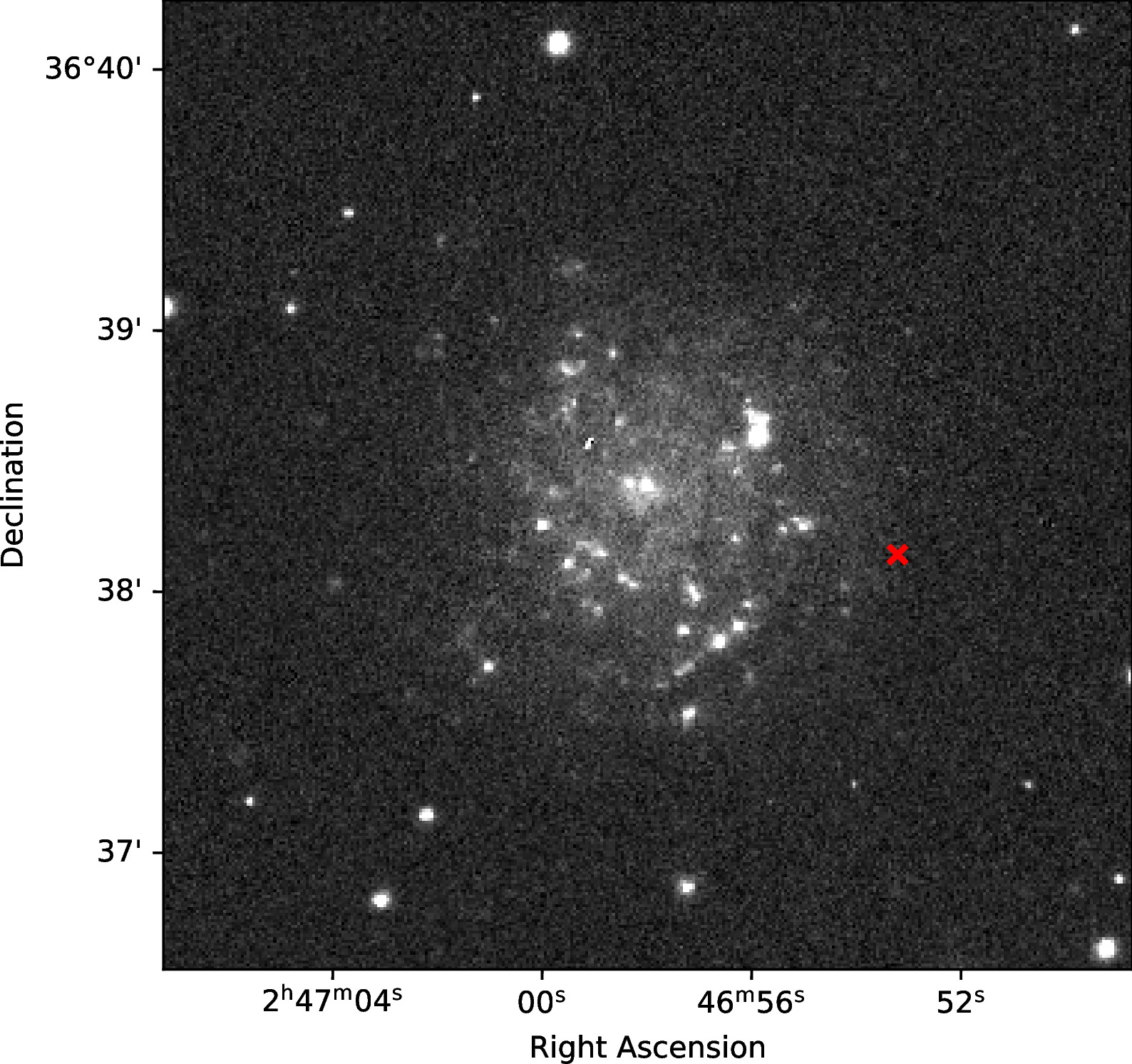}
\includegraphics[width=0.4\textwidth]{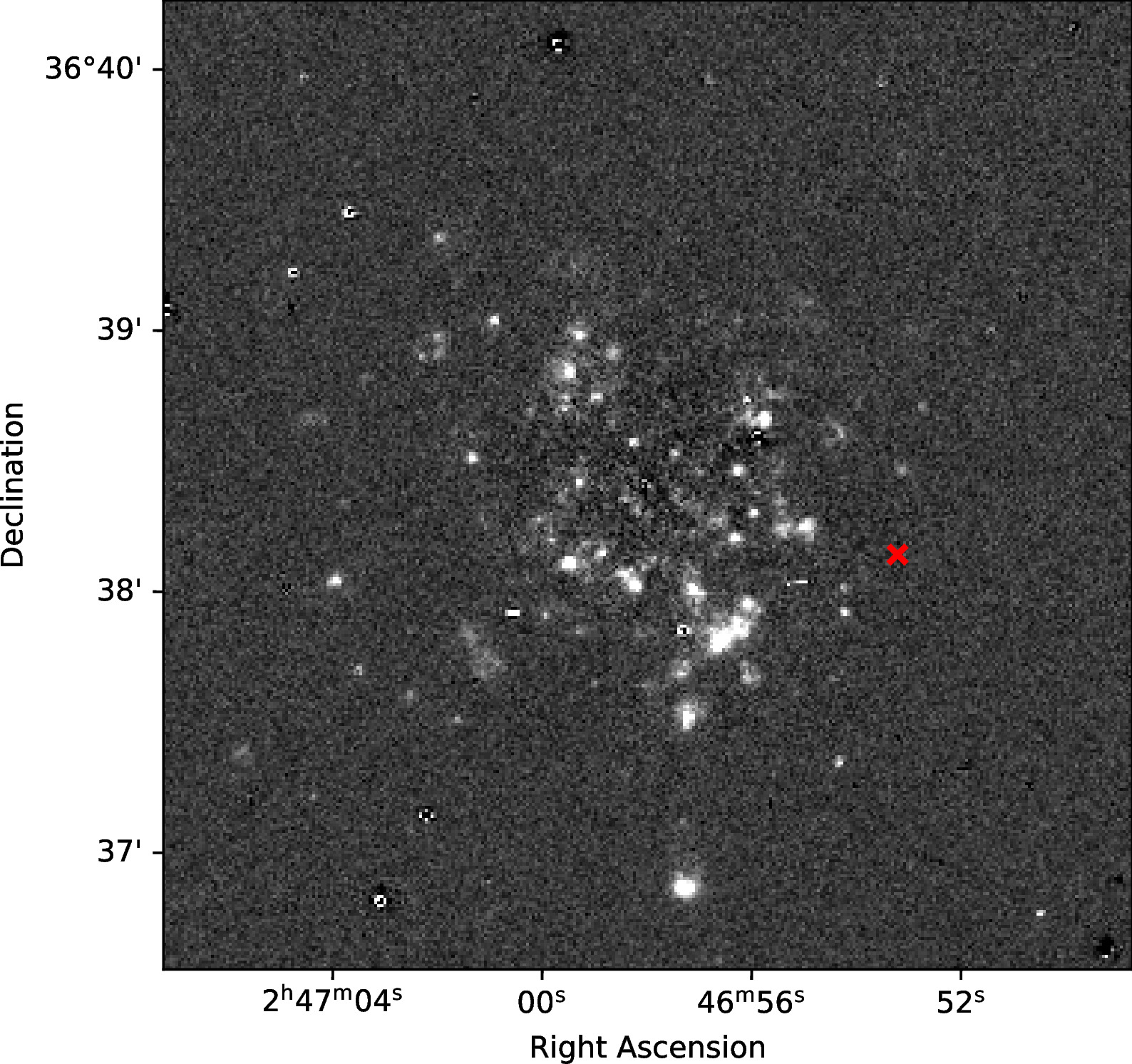}
\includegraphics[width=0.4\textwidth]{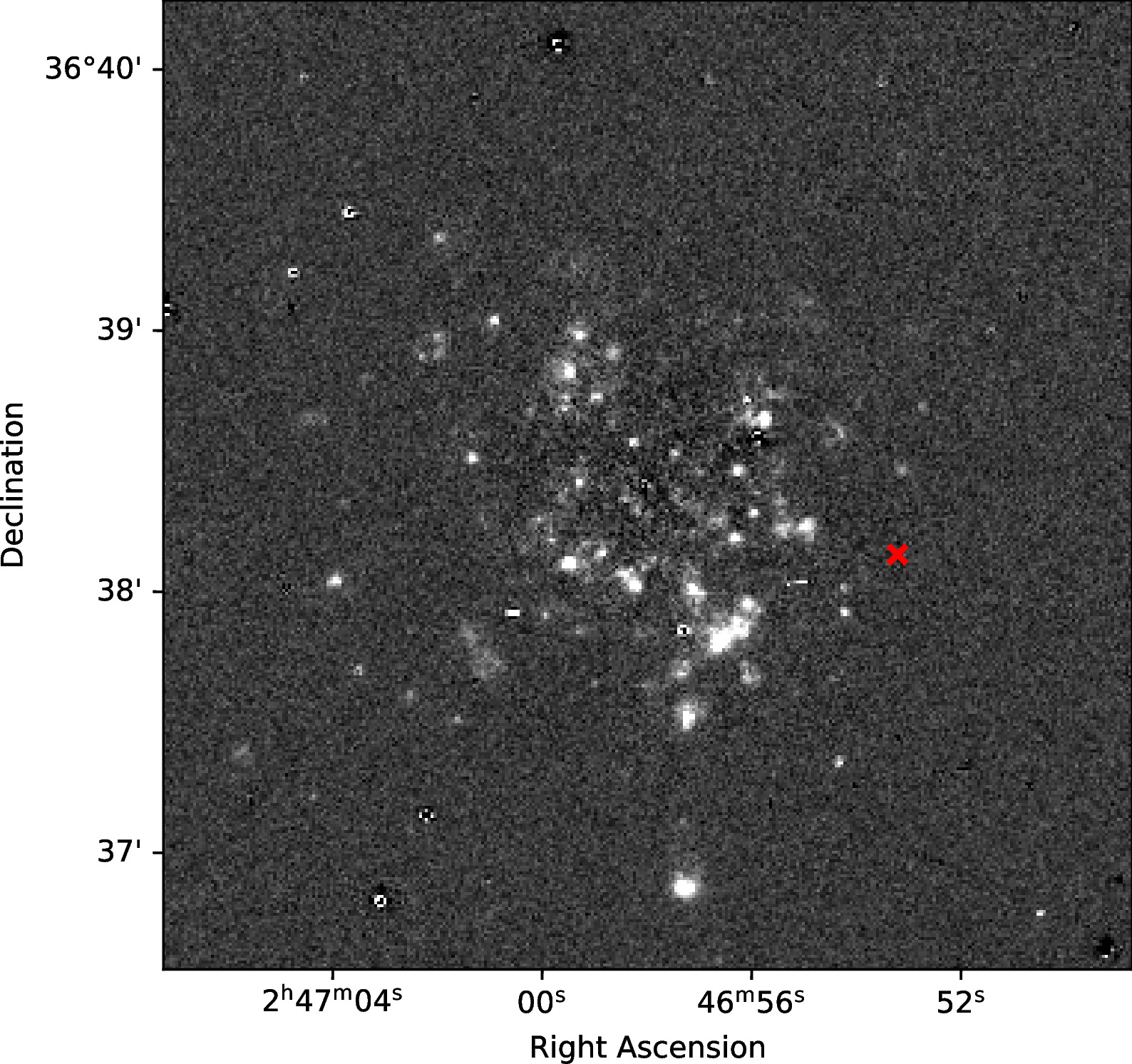}
\caption{J-PLUS NGC1058 $J0660$ image with the type II SN1961V marked in red. Top left: Before dust extinction correction. Top right: After dust extinction correction. Bottom left: After continuum subtraction. Bottom right: After [NII] removal.}
\label{corrections}
\end{figure*}

\subsection{Dust extinction correction}
The Milky Way interstellar dust reddening was corrected using the python code {\sc dustmaps} \citep{2018JOSS....3..695M} and the \textit{Gaia Total Galactic Extinction} map \citep{2023A&A...674A..31D} as the source to obtain the E(B-V) color excess for every galaxy.
In addition, we applied a correction to the reddening due to dust in the SN host galaxy. 
Following \cite{2000ApJ...533..682C}, the difference between the intrinsic and observed flux due to extinction is given by
\begin{equation}
    F_i(\lambda)=F_0(\lambda)10^{0.4E(B-V)k'(\lambda)}
    \label{fluxextintion},
\end{equation}
where $F_i(\lambda)$ and $F_0(\lambda)$ are the intrinsic and observed fluxes at a wavelength $\lambda$, $E(B-V)$ is the color excess, and $k'(\lambda)$ is the extinction law: an empirical relationship between the amount of extinction and wavelength. We employed the \cite{2000ApJ...533..682C} extinction law, defined as
\begin{equation}\label{extintionlaw}
k'(\lambda) = \left\lbrace
\begin{array}{ll}
2.659(-1.857+1.040/\lambda)+R'_V \\ \qquad\qquad\qquad\qquad\qquad\quad\textup{for }  0.63\,\mu m \leq \lambda \leq 2.20 \,\mu m\\ \\
2.659(-2.156+1.509/\lambda-0.198/\lambda^2+0.011/\lambda^3) +R'_V \\ \qquad\qquad\qquad\qquad\qquad\quad\textup{for }  0.12\,\mu m \leq \lambda \leq 0.63 \,\mu m,
\end{array}
\right.  
,\end{equation}
where $R'_V$ is the obscuration in the  $V$ band. Here, we used the same value found by \cite{2000ApJ...533..682C}, $R'_V=4.05\pm0.80,$ for stellar extinction\footnote{The stellar extinction mainly affects those filters without emission lines, and gas extinction mainly affects those with emission lines \citep{Thomas}.}.
To estimate the color excess $E(B-V)$, we followed \cite{dustandNII}, who found a relation between the observed $g'-i'$ color and the spectroscopically measured $E(B-V)$ obtained after convolving the
SDSS spectra with the J-PLUS photometric system. 
These authors obtained the following expression by fitting a power-law function:
\begin{equation}
    E(B-V)=0.206(g'-i')^{1.68}-0.0457,
    \label{colorexcexpresion}
\end{equation}
and assuming $E(B-V)=0$ for $g'-i'<0.4$.

This is an empirical expression for the gas extinction. \cite{dustandNII} mixed in the same process of correction to the stellar and gas components for all the J-PLUS data; so, we used the same method to be coherent. The gas extinction is, at minimum, the same as the stellar extinction. However, on average, the extinction in HII regions can be twice as high \citep{2005MNRAS.358..363C}. Therefore, it is essential to note that the extinction correction applied by this method represents a conservative lower bound.

With all these ingredients, we corrected the fluxes of the 12 images for all SNe in our sample, pixel by pixel. This improvement results in a change in our NCR values compared to studies that do not apply this correction to the fluxes. Hence, we caution against making direct comparisons of NCR values to previous works that employed different methodologies.

\subsection{J0660 continuum subtraction}

J-PLUS narrow-band filters were designed to cover wavelength regions of a number of emission lines of interest. However, they also include the flux of the underlying stellar continuum that needs to measure the gas-phase component exclusively.
In particular, both the $rSDSS$ and $J0660$ filters are collecting the flux of the H$\alpha$ emission line, so to study the H$\alpha$ emission, which is a tracer of the star formation, we have to subtract the stellar continuum from the $J0660$ flux. 

\cite{dustandNII} presented two methods to remove the underlying continuum from a narrow-band filter. 
The simplest is by combining the narrow band that contains the emission with an adjacent or overlapping broad filter that traces the continuum. 
For H$\alpha$, we can use the $J0660$ and the $rSDSS$ filters, but taking  
into account that $J0660$ also includes the contribution of the forbidden transition of [\ion{N}{ii}] doublet (at 654.8 nm and 658.53 nm).
Assuming a flat continuum, the flux of the three emission lines can be recovered using the following expression:
\begin{equation}
    F_{H\alpha+[N II]}=\Delta_{J0600}\frac{\overline{F}_{J0600}-\overline{F}_{rSDSS}}{1-\frac{\Delta_{J0600}}{\Delta_{rSDSS}}},
    \label{2filt}
\end{equation}
where $\overline{F}_{J0600}$ and $\overline{F}_{rSDSS}$ correspond to the average flux inside the $J0660$ and $rSDSS$ filters, respectively, and $\Delta_x$ is defined for any pass band x at any wavelength of interest $\lambda_s$ as
\begin{equation}
    \Delta_x\equiv\frac{\int P_x(\lambda)\lambda d\lambda}{P_x(\lambda=\lambda_s)\lambda_s},
\end{equation}
where $P_x$ is the transmission of the pass band x as a function of wavelength. In our case, we took $\lambda_s=\lambda_{H\alpha}$.

This method assumes a linear continuum, so the H$\alpha$ absorption is not taken into account, resulting in an overall bias of approximately 9\% in all the results \citep{dustandNII}. 
When including more filters, such as those available in J-PAS, this method of continuum subtraction improves significantly \citep{2021A&A...647A.158M}.

\subsection{{\rm [\ion{N}{ii}]} removal}


The continuum-subtracted $J0660$ flux includes both the H$\alpha$ and the [\ion{N}{ii}] doublet emission. 
To remove the [\ion{N}{ii}] contribution, we used the bimodal empirical relation between the spectroscopic dust-corrected H$\alpha$ flux and the total H$\alpha$+[\ion{N}{ii}] flux found by \cite{dustandNII}.
This bimodality can be disentangled by using the same $g'-i'$ color used for the dust correction. 
\cite{dustandNII} obtained the following expressions by fitting a line to each branch:
\begin{equation}\label{niicorrection}
log(F_{H\alpha}) = \left\lbrace
\begin{array}{ll}
0.989log(F_{H\alpha+[N II]})-0.193, & \textup{if }  g'-i'\leq0.5\\
0.954log(F_{H\alpha+[N II]})-0.753, & \textup{if }  g'-i'>0.5
\end{array}
\right.
.\end{equation}
In addition, we can easily obtain the [\ion{N}{ii}] flux by simply taking the difference between the H$\alpha$ flux and $F_{H\alpha+[N II]}$,
\begin{equation}
    F_{[N II]}=F_{H\alpha+[N II]}-F_{H\alpha}.
    \label{niiflux}
\end{equation}

We did not apply this procedure for SNe at $z>0.014$, since the [\ion{N}{ii}] $\lambda$6583 line is outside the J0660 filter range at a higher redshift and a subtraction of this line cannot be correctly performed. Therefore, we assume all emission in the J0660 filter comes from H$\alpha$ flux for $z>0.014$.

\subsection{Other continuum subtractions}

We repeated the subtraction procedure for other narrow filters by applying a different method that was presented in \cite{3filters} and \cite{dustandNII}, in which three instead of two filters were used, one narrow filter that contain the feature of interest and two filters (one at each side if the main feature) to trace the continuum,
\begin{equation}
    F_{cs}=\frac{(\overline{F}_{B1}-\overline{F}_{B2})-\left(\frac{\alpha_{B1}-\alpha_{B2}}{\alpha_{N}-\alpha_{B2}}\right)(\overline{F}_{N}-\overline{F}_{B2})}{\beta_{B1}-\beta_N\left(\frac{\alpha_{B1}-\alpha_{B2}}{\alpha_{N}-\alpha_{B2}}\right)},
    \label{3filteq}
\end{equation}
where $F_{cs}$ is the flux of the narrow filter with the continuum subtracted, $\overline{F}_x$ is the average flux, and $\beta$ and $\alpha$\footnote{We need to make a calculation that involves an integration of the transmission curve of the pass band; the data of the transmission curve are the same and can be found on the J-PLUS webpage.} are defined as
\begin{equation}
    \beta_x\equiv\frac{P_x(\lambda=\lambda_s)\lambda_s}{\int P_x(\lambda)\lambda d\lambda} \quad ; \quad \alpha_x\equiv\frac{\int P_x(\lambda)\lambda^2 d\lambda}{\int P_x(\lambda)\lambda d\lambda}.
\end{equation}
The subindices $B1$, $B2,$ and $N$ refer to the first broad filter, the second one, and the narrow filter, respectively. We note that in this case the continuum is not flat, but it has a slope defined by the two reference filters. Similarly to the method of subtracting the continuum described above, this method also improves significantly when additional filters are incorporated \citep{2021A&A...647A.158M}.

We used this method to subtract the continuum of the $J0378$ and $J0861$ filter using the $uJAVA$ and $gSDSS$ and the $iSDSS$ and $zSDSS$ filters, respectively. In the latter case, this is an estimate of the calcium triplet absorption. More negative NCR values for absorption filters indicate deeper absorption. Therefore, we inverted the sign of the resulting fluxes when constructing the NCR with this continuum-subtraction method to maintain coherence during analysis. 

\subsection{Error calculation}

Throughout the preceding calculations, it is essential to consider the respective error estimation through robust propagation.
The total uncertainty in our measurements of the flux (Eq. \ref{fluxec}) for a source and a given filter (\citealt{errMolino}, \citealt{error2}) is given by
\begin{equation}
    \sigma_{F_{\lambda}}=\sqrt{\left(\frac{\partial F_{\lambda}}{\partial ZP}\sigma_{ZP}\right)^2+\left(\frac{\partial F_{\lambda}}{\partial C}\sigma_{bn}\right)^2+\left(\frac{\partial F_{\lambda}}{\partial C}\sigma_{ec}\right)^2},
    \label{errorflux}
\end{equation}
$\sigma_{ZP}$ being the error of the zero point reflected in Table \ref{Tzp}, $\sigma_{ec}$ the uncertainly in the electron counting of the CCD, and $\sigma_{bn}$ the large-scale background noise variation; the last two are given by
\begin{equation}
    \sigma_{ec}=\sqrt{\frac{C}{G}}, \quad \sigma_{bn}=S_{fit}\sqrt{N_{pix}}\left(a_{fit}+b_{fit}\sqrt{N_{pix}}\right),
\end{equation}
where G is the gain of the detector, $N_{pix}$ the number of pixels; and $S_{fit}$, $a_{fit,}$ and $b_{fit}$ are the resulting coefficients
from the fitting. This information is found within headers of each image.

We calculated the error of the other magnitudes by a quadratic propagation of errors.
The error of the flux with the dust correction (Eq. \ref{fluxextintion}) is given by
\begin{equation}
    \sigma_{F_i(\lambda)}= \sqrt{(D_{F_0}\cdot\sigma_{F_0(\lambda)})^2+(D_{k'}\cdot\sigma_{k'(\lambda)})^2+(D_{E}\cdot\sigma_{E(B-V)})^2},
    \label{fluxcorrecterr}
\end{equation}
where $D_{F_0}=\frac{\partial F_i(\lambda)}{\partial F_0(\lambda)}$, $D_{k'}=\frac{\partial F_i(\lambda)}{\partial k'(\lambda)}$, and $D_{E}=\frac{\partial F_i(\lambda)}{\partial E(B-V)}$.

\noindent $\sigma_{E(B-V)}$ and $\sigma_{k'(\lambda)}$ are the errors of the color excess (Eq. \ref{colorexcexpresion}) and the error given for the extinction law (Eq. \ref{extintionlaw}), respectively, that are given by
\begin{equation}
    \sigma_{E(B-V)}=\sqrt{\left(\frac{\partial E(B-V)}{\partial i'}\sigma_{i'}\right)^2+\left(\frac{\partial E(B-V)}{\partial g'}\sigma_{g'}\right)^2},
\end{equation}
where $\sigma_{g'}$ and $\sigma_{i'}$ are the error of the flux in the $g_{SDSS}$ and $i_{SDSS}$ bands given by Equation \ref{errorflux}, and 
\begin{equation}
    \sigma_{k'(\lambda)}=\sqrt{\left(\frac{\partial k'(\lambda)}{\partial \lambda}\sigma_{\lambda}\right)^2+\left(\frac{\partial k'(\lambda)}{\partial R'_V}\sigma_{R'_V}\right)^2}.
\end{equation}
Knowing the error of the fluxes with the dust correction, we can now calculate the error of the continuum subtracted fluxes.
In the case of the subtraction of the red continuum of the $J0660$ filter (Eq. \ref{2filt}), we have

\begin{equation}
    \sigma_{F_{H\alpha+[NII]}}=\sqrt{\left(\frac{\partial F_{H\alpha+[NII]}}{\partial \overline{F}_{F600}}\sigma_{\overline{F}_{F600}}\right)^2+\left(\frac{\partial F_{H\alpha+[NII]}}{\partial \overline{F}_{r'}}\sigma_{\overline{F}_{r'}}\right)^2},
\end{equation}
with $\sigma_{\overline{F}_{r'}}$ and $\sigma_{\overline{F}_{F600}}$ being the error of the extinction-corrected flux of the $J0660$ and $r_{SDSS}$ filters obtained by Equation \ref{fluxcorrecterr}.

\begin{figure}
                \includegraphics[width=0.5\textwidth]{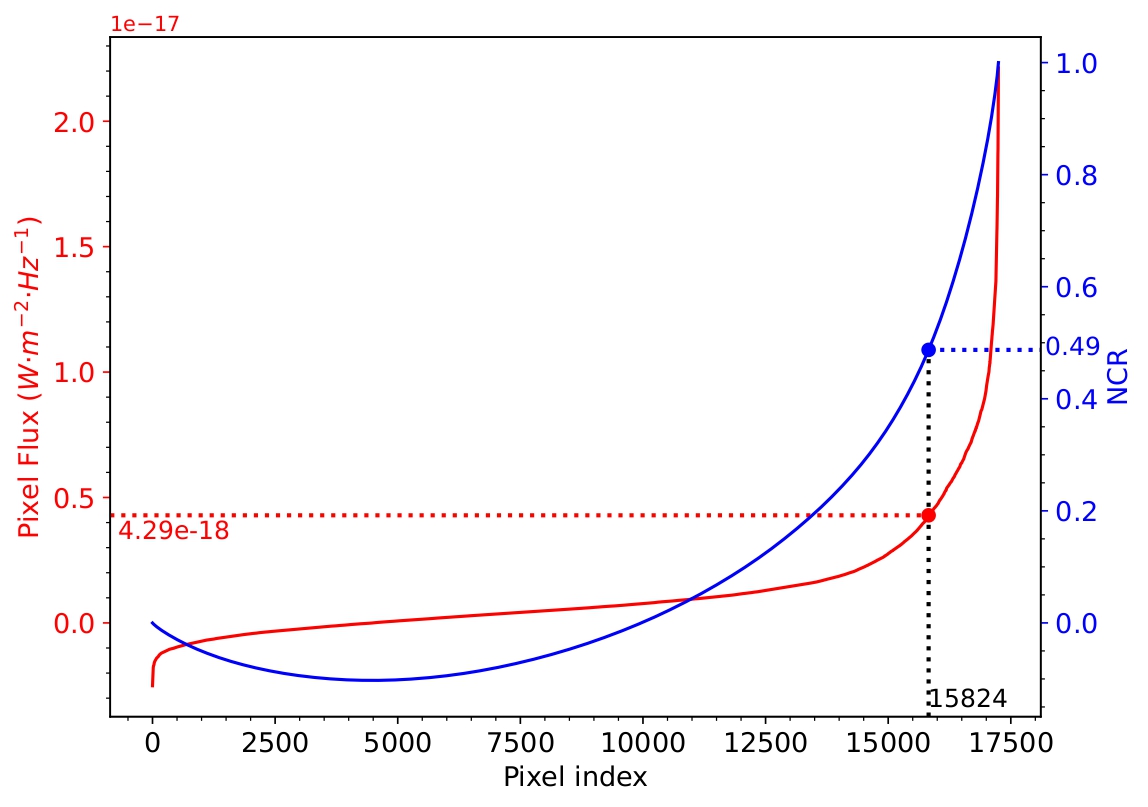}
        \caption{Example of calculation of NCR value for SN pixel. The red curve is the sorted pixel value (left flux scale), while the blue curve is the cumulative distribution of the flux values, i.e., the normalized cumulative-rank pixel-value function (NCRPVF or NCR for simplicity).
        The blue dot represents the position and corresponding NCR value of the SN 2008az pixel, and the red dot represents the corresponding flux value.
}
        \label{cumulativecurves}
\end{figure}

We can then calculate the error of the H$\alpha$ flux with the [NII] line subtracted (Eq. \ref{niicorrection}) and the error of the [NII] line flux (Eq. \ref{niiflux}) as
\begin{equation}
    \sigma_{F_{H\alpha}}=\frac{\partial F_{H\alpha}}{\partial_{F_{H\alpha+[NII]}}}\sigma_{F_{H\alpha+[NII]}}
\end{equation}
and
\begin{equation}
    \sigma_{F_{[NII]}}=\sqrt{\left(\frac{\partial F_{[NII]}}{\partial F_{H\alpha}}\sigma_{F_{H\alpha}}\right)^2+\left(\frac{\partial F_{[NII]}}{\partial{F_{H\alpha+[NII]}}}\sigma_{F_{H\alpha+[NII]}}\right)^2}.
\end{equation}

Finally, we can calculate the error of the $J0378$ and $J0861$ flux with the continuum subtracted (Eq. \ref{3filteq}) as
\begin{equation}
    \sigma_{F_{cs}}=\sqrt{\left(\frac{\partial F_{cs}}{\partial \overline{F}_{N}}\sigma_{\overline{F}_{N}}\right)^2+\left(\frac{\partial F_{cs}}{\partial \overline{F}_{B1}}\sigma_{\overline{F}_{B1}}\right)^2+\left(\frac{\partial F_{cs}}{\partial \overline{F}_{B2}}\sigma_{\overline{F}_{B2}}\right)^2},
\end{equation}
with the subindices B1, B2, and N being referred to the first broad filter, the second one, and the narrow filter, respectively.
Errors on $\alpha_x$, $\beta_x,$ and $\Delta_x$ are negligible and are therefore not included in the calculation in order to simplify the procedure.

\subsection{Final 3D data cubes}

Once all these corrections are applied, we store these newly generated images as new slices in the 3D data cube, where fluxes are in the primary extension and flux errors are in the first extension. The final cube has 17 slices: 12 corresponding to the extinction-corrected images in the 12 J-PLUS filters, the H$\alpha$+[\ion{N}{ii}] continuum subtracted flux, the $J0378$ and $J0861$ continuum-subtracted flux, the flux corresponding only to the H$\alpha$ line, and the [\ion{N}{ii}] emission line flux. These resulting data cubes are available at {\sc zenodo}\footnote{\href{https://zenodo.org/doi/10.5281/zenodo.10514631}{https://zenodo.org/doi/10.5281/zenodo.10514631}} for all 418 SNe host galaxies presented in this work.
Figure \ref{corrections} shows the H$\alpha$ image of the same galaxy from Figure \ref{frame} at four different steps of the analysis: (a) the initial cutout from the observed image; (b) after dust reddening correction; (c) after $r$-band continuum subtraction; and (d) after [\ion{N}{ii}] emission  removal.

\begin{figure*}
\includegraphics[trim=0cm 0cm 0cm 0cm, clip=True,width=0.31\textwidth]{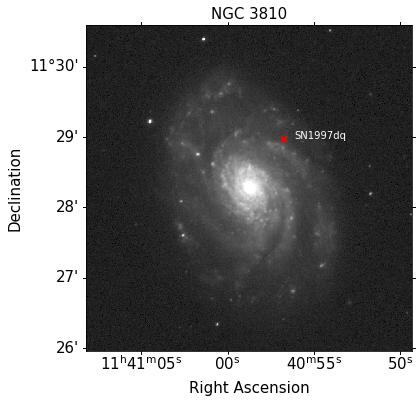}
\includegraphics[trim=0cm 0cm 0cm 0cm, clip=True,width=0.65\textwidth]{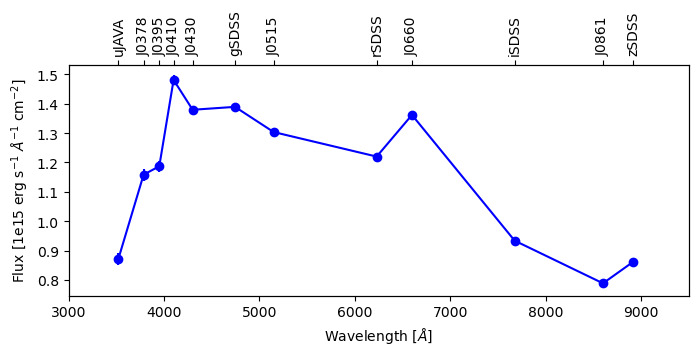}
\includegraphics[trim=0cm 0cm 0cm 0cm, clip=True,width=0.31\textwidth]{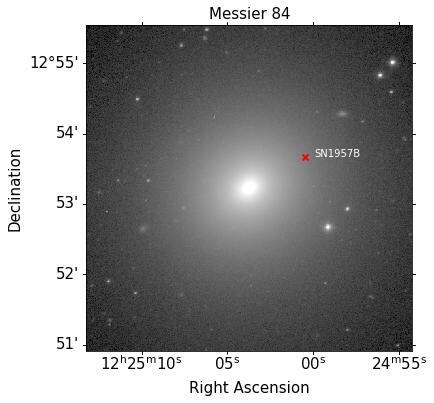}
\includegraphics[trim=0cm 0cm 0cm 0cm, clip=True,width=0.65\textwidth]{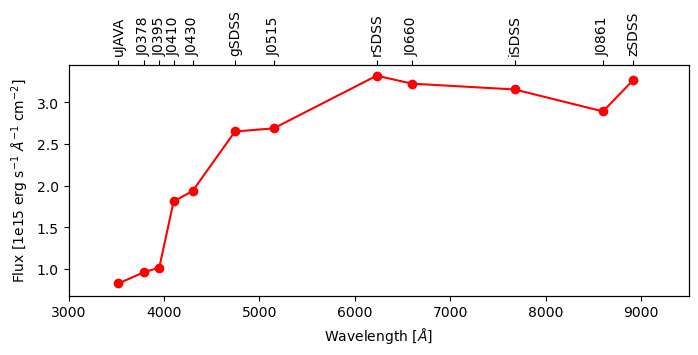}
\caption{Example of two SEDs of SN local environments of one squared kiloparsec extracted from J-PLUS observations. On top, the star-forming environment of SN~1997dq in its host galaxy NGC~3810 clearly shows a blue continuum based on the five broad-band filters and increased flux for the narrow-band filters where the main emission lines fall. On the bottom, for SN~1957B in the Messier 84 galaxy, a smooth red continuum is seen with no bumps at the narrow-band filter wavelengths.
}
\label{sed}
\end{figure*}


\section{Analysis}

\subsection{Normalized cumulative-rank distributions}

The NCR is obtained by sorting the flux values in increasing order (see Figure \ref{cumulativecurves}), constructing the cumulative distribution, and normalizing it to the total emission of the galaxy. This procedure associates each pixel with an NCR value between 0 and 1, where the brightest pixel in the galaxy has an NCR value of one, and all pixels corresponding to sky have an NCR value of zero. Then, the ranked value in this distribution corresponding to the pixel where the SN occurred\footnote{The precision of SN locations is lower than the image resolution (see Section 1), so what we are measuring is the immediate SNe environment.} is the NCR associated with the SN.
By compiling significant numbers of NCR values for different SN types, we can build NCR distributions and study the differences among types and use this information to infer properties of their progenitors.

The NCR method has previously been applied to H$\alpha$ imaging of SN hosts. Assuming that the H$\alpha$ emission scales by the number of stars that are formed \citep{1998ApJ...498..541K}, a diagonal cumulative NCR distribution with a mean value of 0.5 indicates that the population traces the observed light, and the SN type follows the number of stars formed and mapped by that particular SF tracer. Therefore, if SNe explode predominantly in locations with high SFR, they will favor higher NCRs. On the other hand, SNe that explode in random locations over the galaxy favor low NCRs.

%

%
%
Following \cite{2006A&A...453...57J}, we included all negative flux values in the construction of the NCR function; such values are the result
of a background subtraction during the processing with the J-PLUS pipeline \citep{2019A&A...622A.176C}. 
This NCR calculation is applied independently to the 17 images of the 418 SN host galaxies.

\subsection{SN environment SED fitting}

To extract the main properties of SN environments, we construct SEDs by measuring aperture photometry in all 12 J-PLUS bands in circular apertures of 1 kpc$^2$ centered at SN locations. 
To ensure the analysis yields meaningful results, the fluxes need to be only corrected for Milky Way extinction, so here we use the constructed 3D cubes with only this correction applied (at the stage outlined in Section 3.3).
The size of the aperture is selected so that the flux contained in the aperture has a high enough signal-to-noise ratio to reliably obtain SSP parameters. We made calculations using the host galaxy redshift obtained from the OSC and assuming a flat $\Lambda$CDM cosmology with $H_0=70$ km s$^{-1}$ Mpc$^{-1}$. 
Within the redshift range of our sample this translates into apertures of a 26 to 2 arcsec radius.
Figure \ref{sed} shows a couple of examples of the $rSDSS$ filter galaxy cutout and the SED from the 12 J-PLUS bands. We note how the SED shape of an SN environment with ongoing star formation (SN 1997dq on top) shows a bluer continuum with a clear H$\alpha$ emission, while the SED of a SN environment in a passive galaxy (SN 1957B on the bottom) looks redder and smoother.

\begin{figure*}
\centering
\includegraphics[trim=0.2cm 0.3cm 0.2cm 0.4cm, clip=True, width=0.32\textwidth]{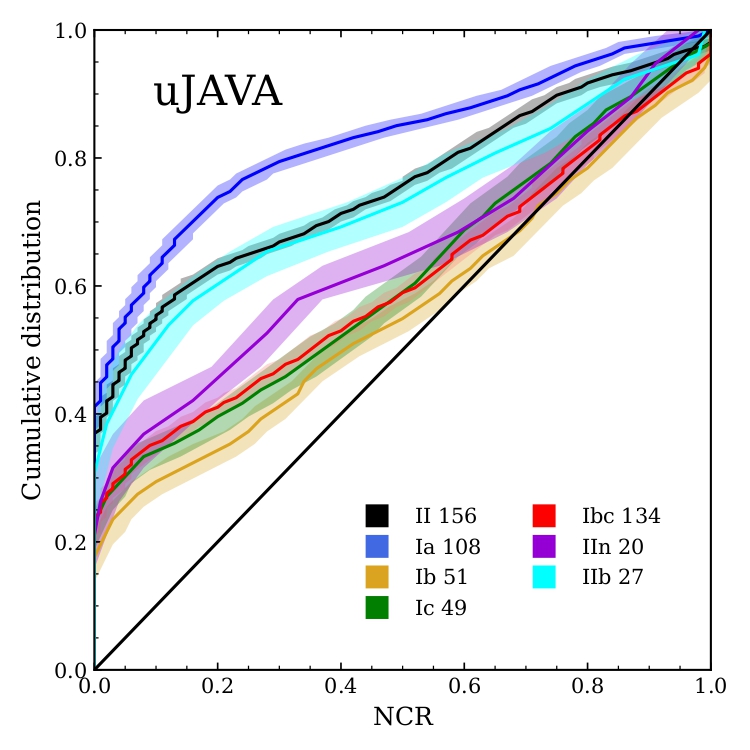}
\includegraphics[trim=0.2cm 0.3cm 0.2cm 0.4cm, clip=True, width=0.32\textwidth]{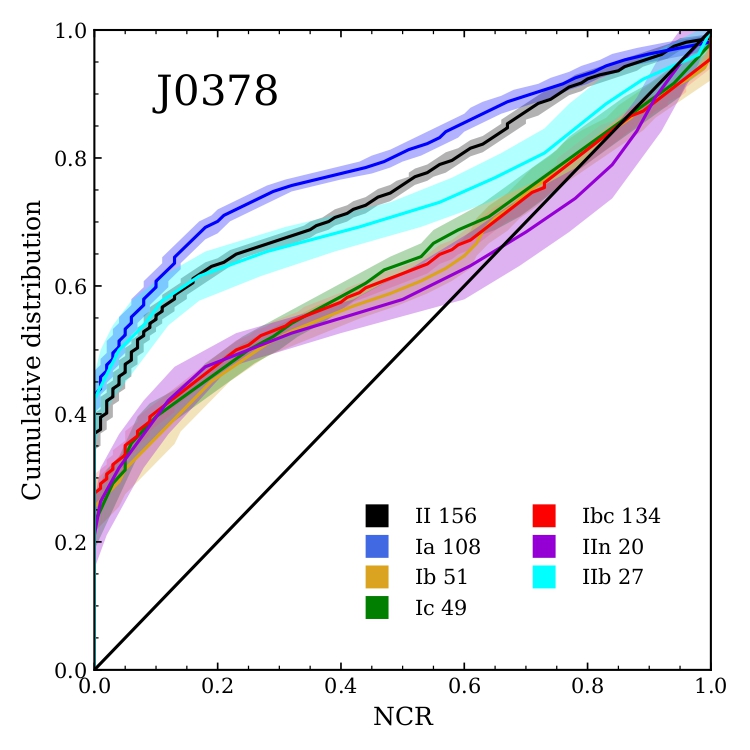}
\includegraphics[trim=0.2cm 0.3cm 0.2cm 0.4cm, clip=True, width=0.32\textwidth]{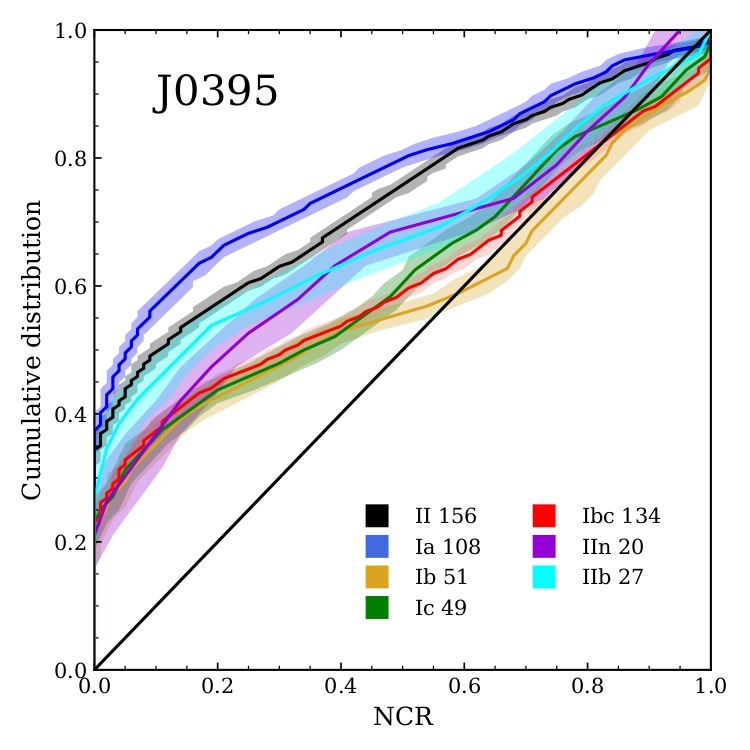}
\includegraphics[trim=0.2cm 0.3cm 0.2cm 0.4cm, clip=True, width=0.32\textwidth]{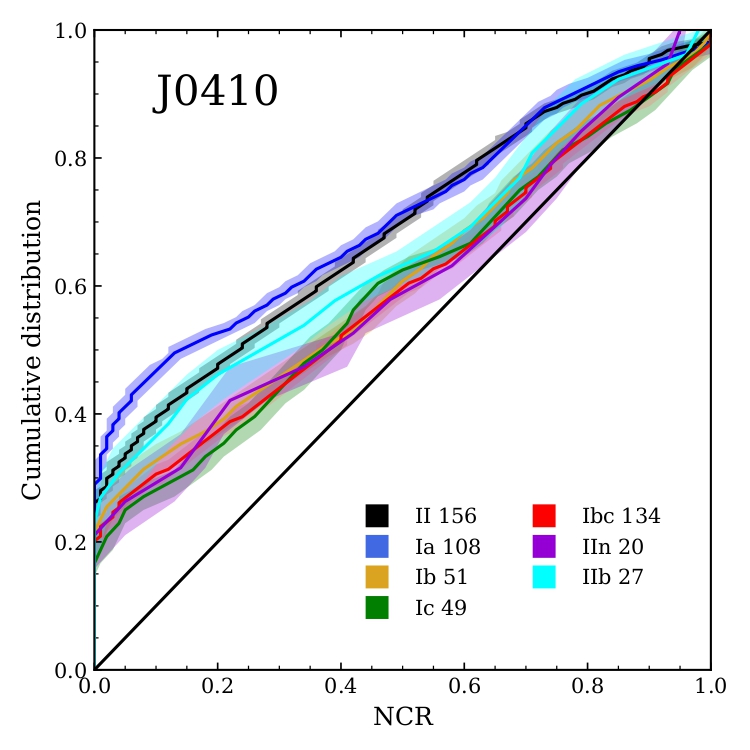}
\includegraphics[trim=0.2cm 0.3cm 0.2cm 0.4cm, clip=True, width=0.32\textwidth]{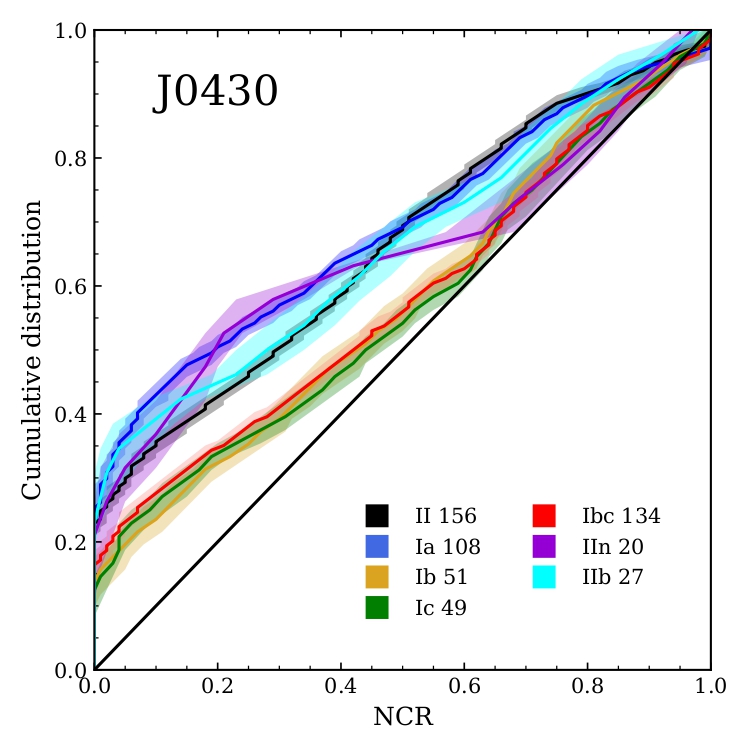}
\includegraphics[trim=0.2cm 0.3cm 0.2cm 0.4cm, clip=True, width=0.32\textwidth]{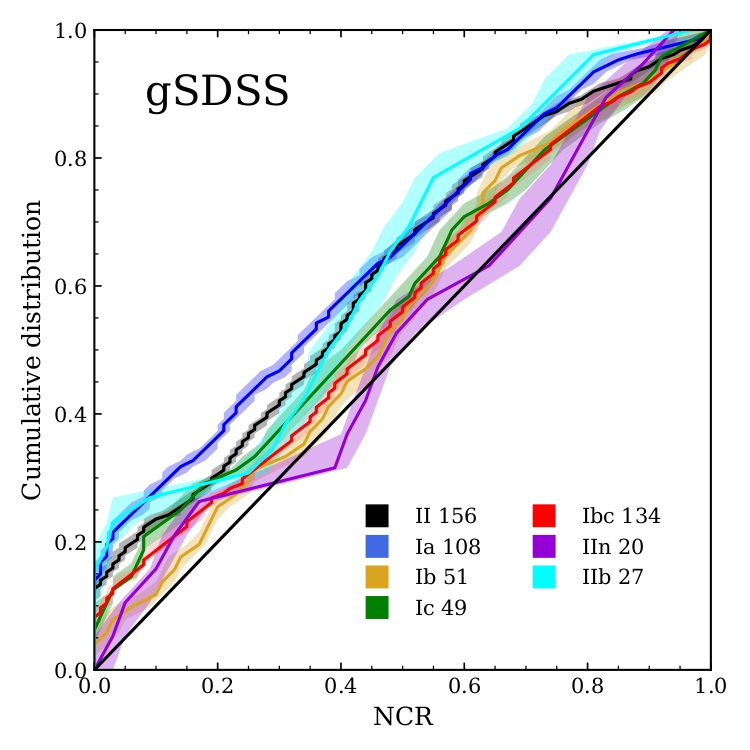}
\includegraphics[trim=0.2cm 0.3cm 0.2cm 0.4cm, clip=True, width=0.32\textwidth]{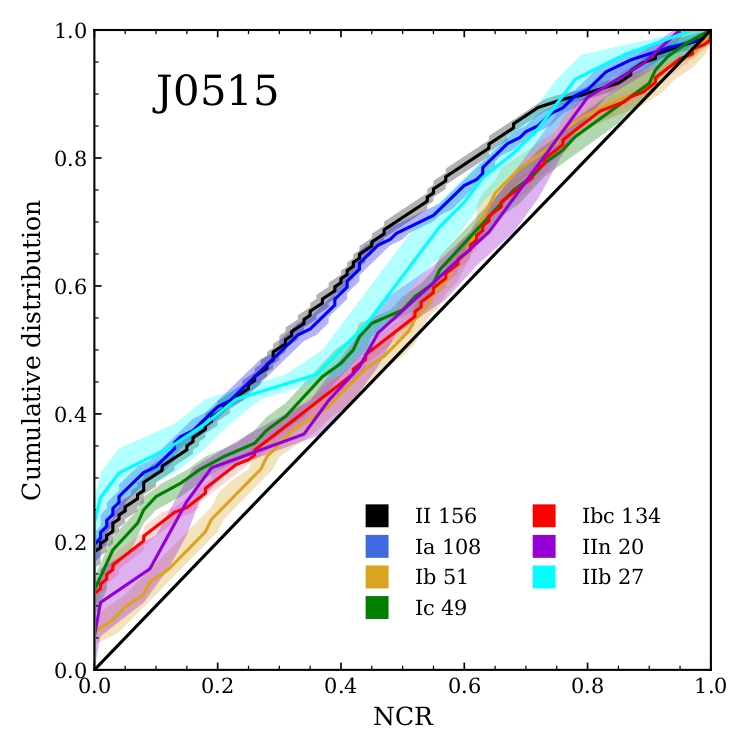}
\includegraphics[trim=0.2cm 0.3cm 0.2cm 0.4cm, clip=True, width=0.32\textwidth]{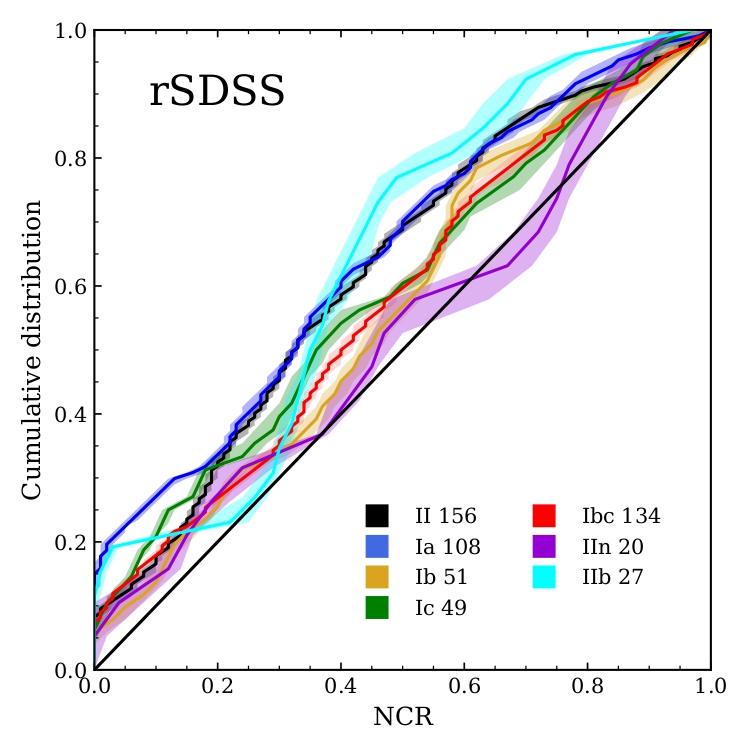}
\includegraphics[trim=0.2cm 0.3cm 0.2cm 0.4cm, clip=True, width=0.32\textwidth]{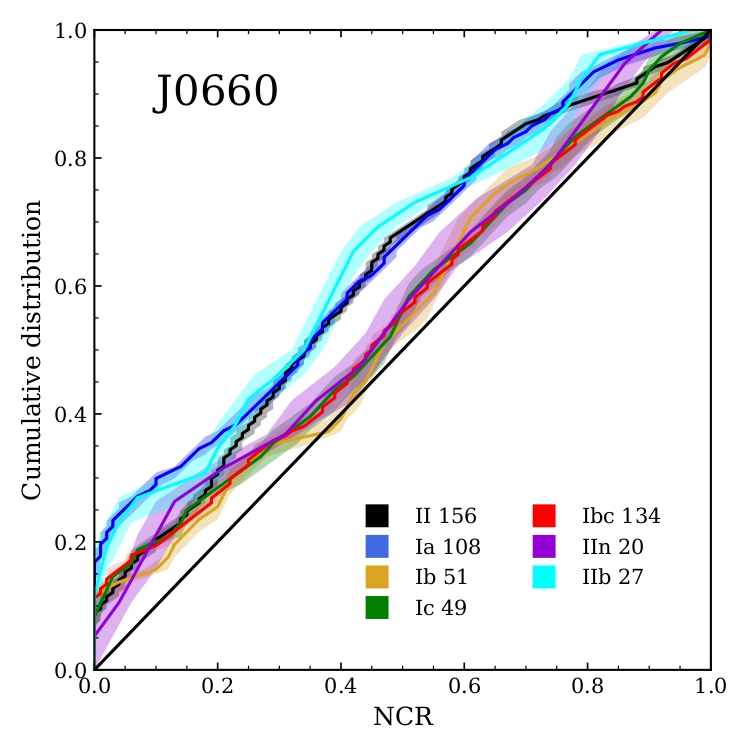}
\includegraphics[trim=0.2cm 0.3cm 0.2cm 0.4cm, clip=True, width=0.32\textwidth]{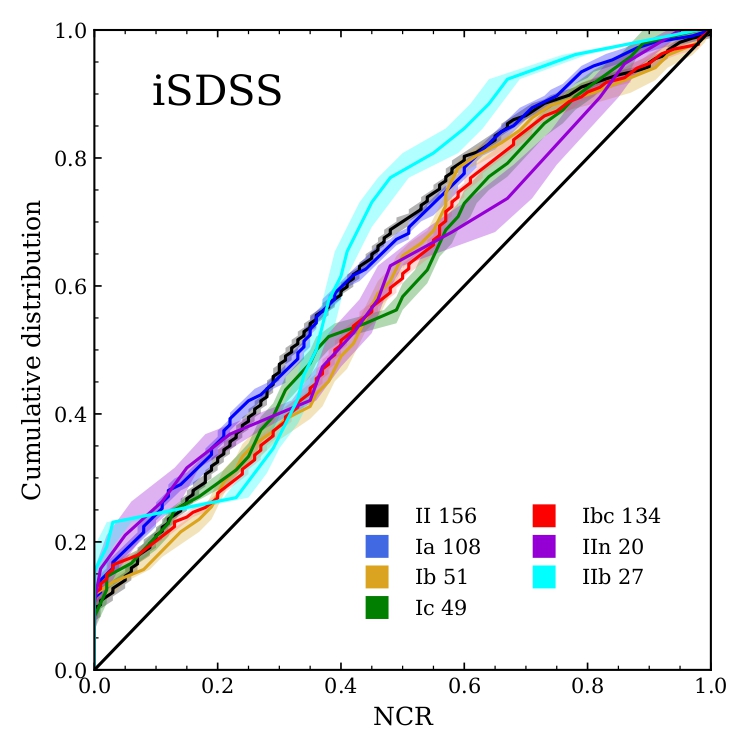}
\includegraphics[trim=0.2cm 0.3cm 0.2cm 0.4cm, clip=True, width=0.32\textwidth]{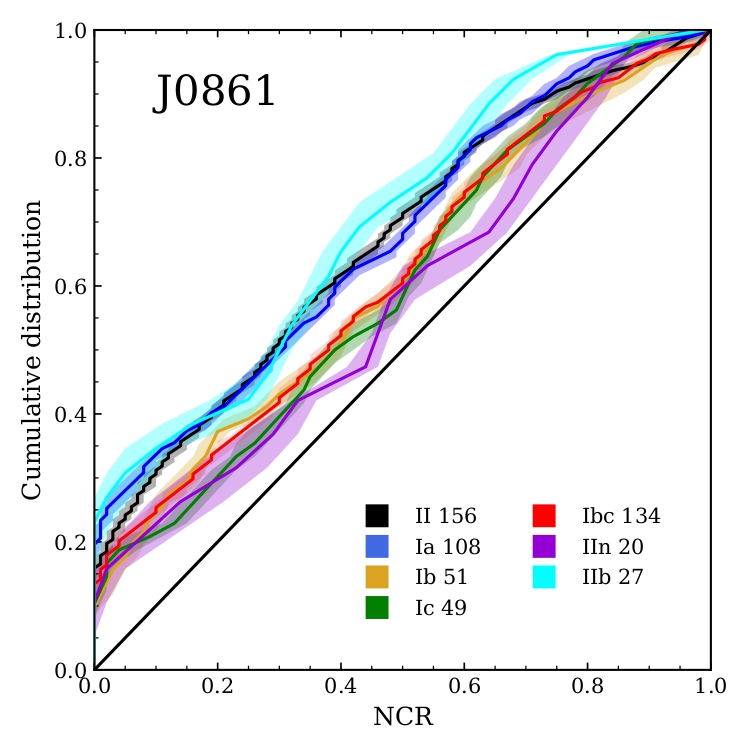}
\includegraphics[trim=0.2cm 0.3cm 0.2cm 0.4cm, clip=True, width=0.32\textwidth]{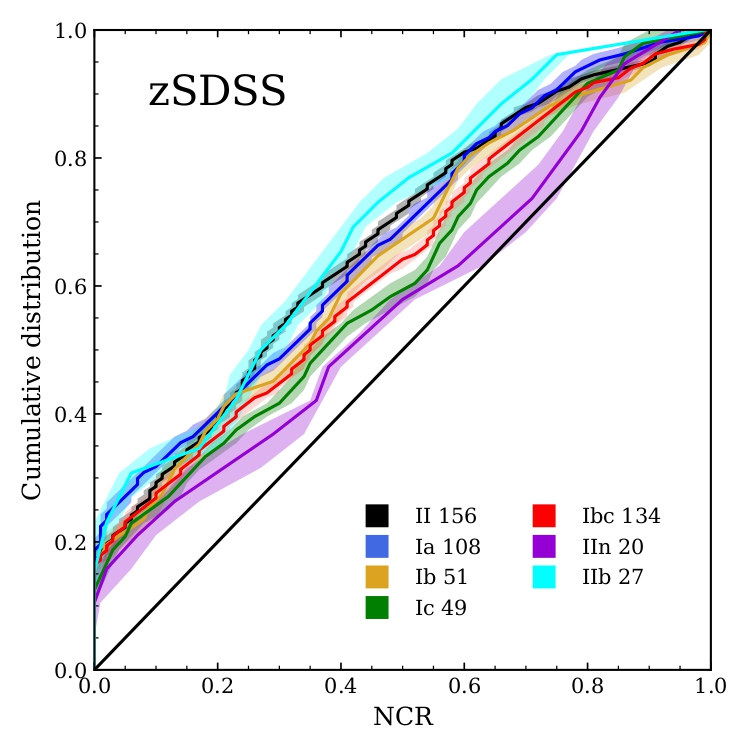}
\caption{Cumulative NCR distributions of 12 broad- and narrow-band filters. The straight black diagonal line represents a hypothetical distribution, infinite in size, which accurately traces the respective observed flux. Each SN distribution corresponds to the median and one sigma of the 30000 NCR trial distributions as detailed in Section 5.1 of \protect\cite{joe2}. It can be observed that as we move to redder filters (from gSDSS to zSDSS), the distributions for each type tend to become more similar along the diagonal. In contrast, there is a clearer distinction between them in the blue filters with more zero values.}
  \label{fig:ncr}       
\end{figure*}

\begin{figure*} 
\centering
\includegraphics[width=0.32\textwidth]{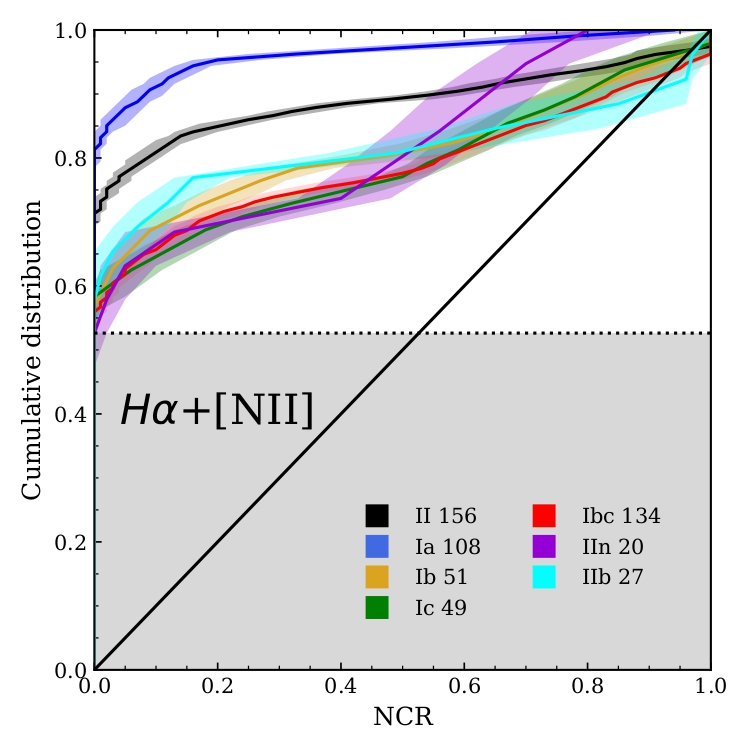}
\includegraphics[width=0.32\textwidth]{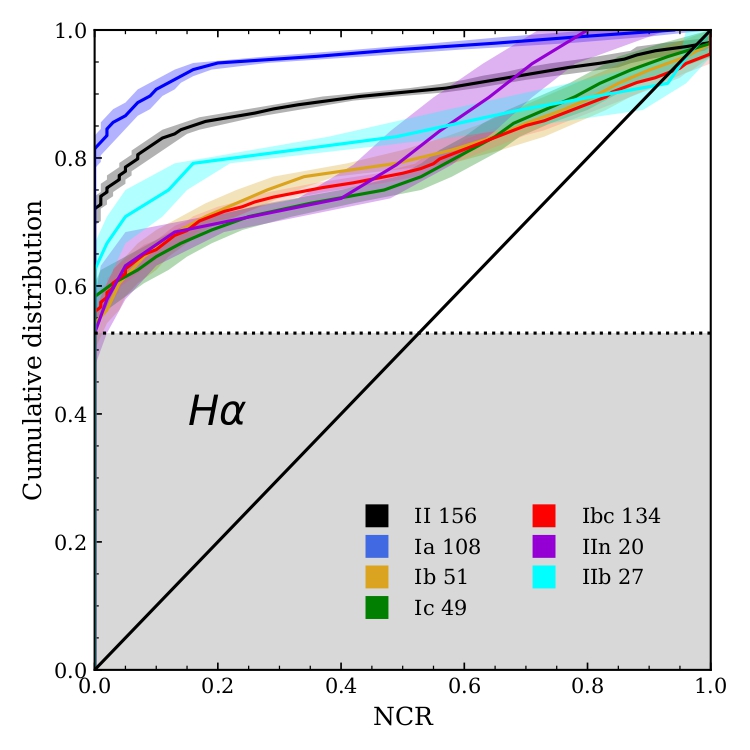}
\includegraphics[width=0.32\textwidth]{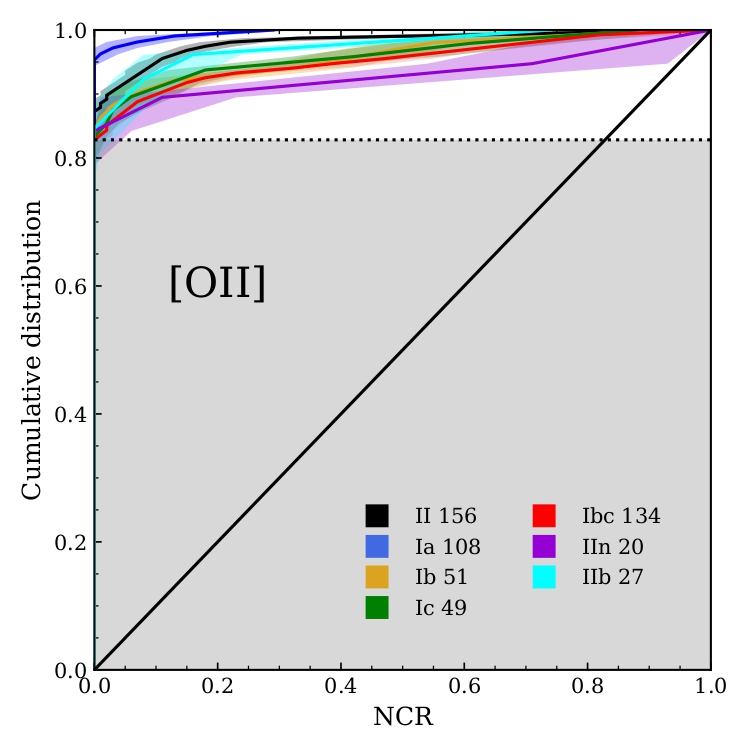}
\includegraphics[width=0.32\textwidth]{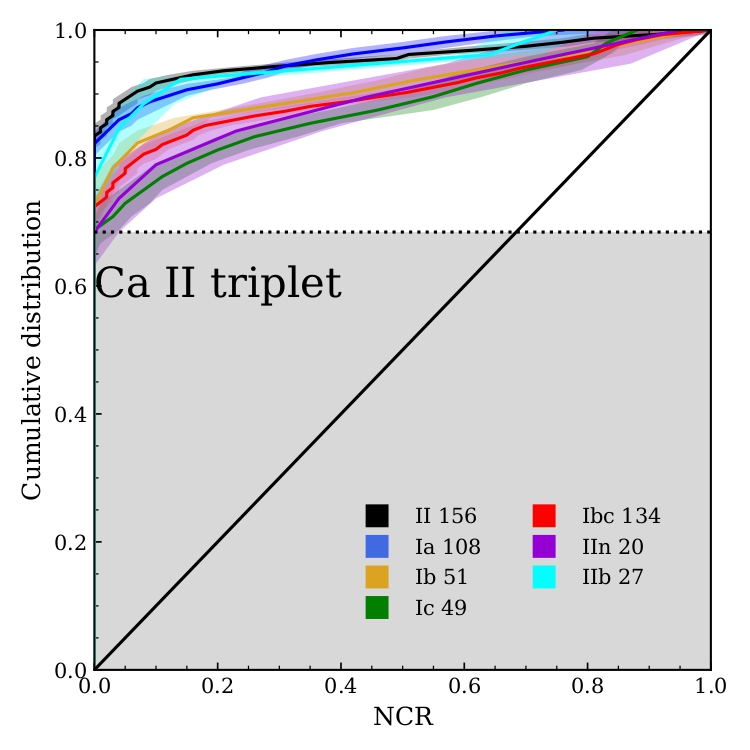}
\includegraphics[width=0.32\textwidth]{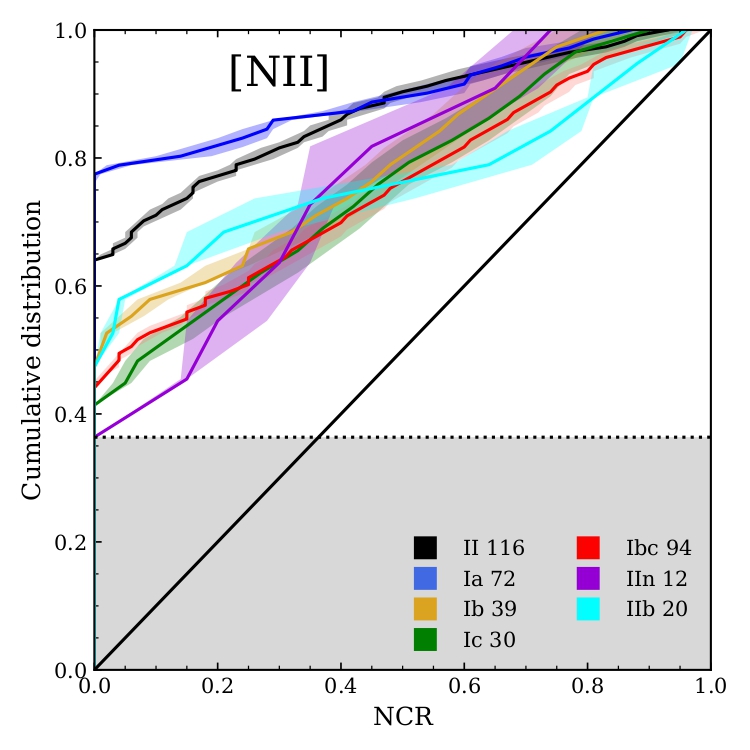}        
\caption{Cumulative NCR distributions of five constructed filters. 
H$\alpha+$[\ion{N}{ii}], [\ion{O}{ii}] and Ca II triplet distributions correspond to the J0660, J0378, and J0861 continuum-subtracted filters, respectively. 
The H$\alpha$ filter is the result of removing the [\ion{N}{ii}] contribution, which is only applied to SNe with $z<0.014$.
All distributions correspond to the median and one sigma of the 30000 NCR trial distributions as outlined in Section 5.1 \citep{joe2}. 
The straight black diagonal line represents a hypothetical distribution, infinite in size, that accurately traces the respective observed flux. 
The gray shading marks the lowest values of zeros in a distribution in each filter. }
\label{fig:ncr2}        
\end{figure*}

Once the 418 SEDs were built, we used FAST++ \citep{2018A&A...618A..85S}\footnote{Available at \href{https://github.com/cschreib/fastpp}{https://github.com/cschreib/fastpp}}, a C++ implementation of FAST \citep{2009ApJ...700..221K}, to fit them and obtain a number of galaxy physical properties.
Given galaxy photometry, FAST++ determines the best-fit SED from a library of simple stellar population (SSP) models.
These models are initially synthesized in composite stellar populations (CSPs) on a 5D grid, with 
each grid point corresponding to a CSP SED with age $t_*$, stellar metallicity $Z_*$, $V$ band extinction $A_V$,  timescale $\tau$, and star formation history (SFH) at a given redshift $z$. At each point on the grid, a $\chi^2$ value is calculated as
\begin{equation}
    \chi^2_F = \sum^{N}_i \frac{\big[F_{\lambda, i} - F_{\lambda, \text{mod}}(t_*, \tau, A_V, Z_*, z)\big]^2}{\sigma^2_i},
\end{equation}
where $N$ is the number of photometric points and $\sigma_i$ is the error for point $i$. One-sigma confidence intervals for parameters and derived quantities (such as stellar mass and SFR) are determined using Monte Carlo sampling of the grid around the lowest $\chi^2$.  
In this work, we used the BC03 \citep{2003MNRAS.344.1000B} SSP library calculated using a delayed exponential (delayed-$\tau$) 
SFH parametrization and a  \cite{2003PASP..115..763C} initial mass function.
Dust extinction is modeled with a  \cite{2000ApJ...533..682C} dust extinction law with a uniform and constant foreground dust screen.
In all fits, the redshift was fixed to the reported spectroscopic redshift of the SN host galaxy.

The output of FAST++ includes the distribution of stellar populations of different ages and metallicities present at that location. 
Although several degeneracies are at play, with large enough samples one can disentangle the underlying differences between the average properties of these parent populations for different SN types. In particular, we mainly focused on stellar population age, dust extinction, stellar mass, and star formation rate.


\section{Results}

\begin{figure*}
\centering
                \includegraphics[width=0.8\textwidth]{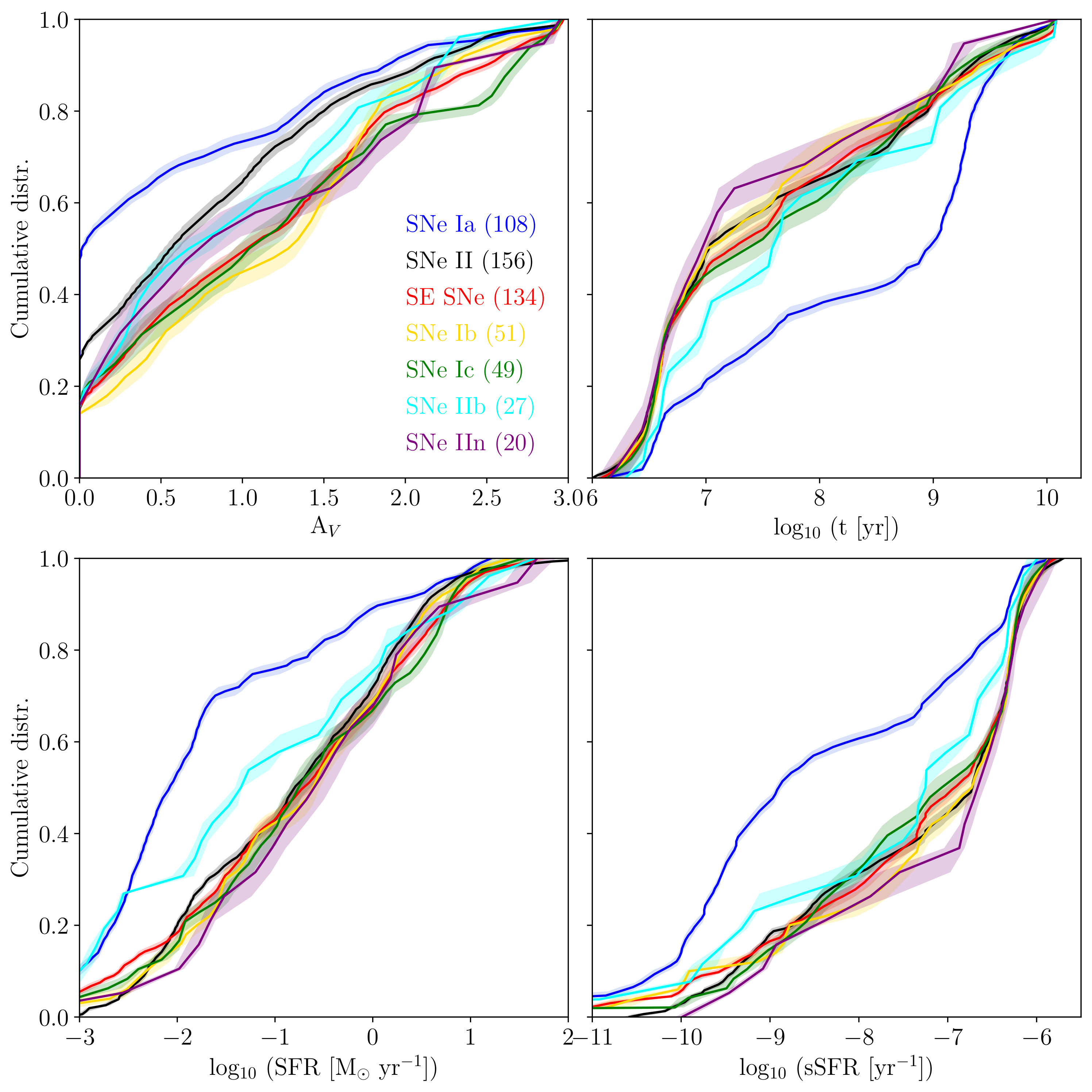}
        \caption{Cumulative distributions of each of three main SN types for main environmental parameters obtained from best FAST++ fits to SED: extinction $A_V$, average stellar age $<t>$, star-formation rate $\log_{10}$ SFR, and specific star-formation rate $\log_{10}$ SFR/Mass. }
        \label{fast++}
\end{figure*}

\subsection{NCR distributions}

The resulting NCRs for all individual SNe are provided in Tables \ref{tab:NCRs_broad}, \ref{tab:NCRs_narrow}, and \ref{tab:NCRs_lines}.
With the NCR and error for all SNe in our sample, we constructed cumulative distributions for all SN types and in all 17 frames of the data cubes as follows. We built a probability distribution for every NCR using 30000 NCR trials of the form $NCR+(\sigma_{NCR}\cdot\alpha$), $\alpha$ being a random variable between -1 and 1. We sorted all the NCRs in increasing order, and the median and one sigma of those distributions of 30000 NCRs are shown in the plots.
In Figure \ref{fig:ncr}, we show the 12 panels, one for each broad and narrow band, with the corresponding NCR distributions for the seven SN types.

In general, we observe that as we move to redder wavelengths (from $gSDSS$ to $zSDSS$), all SN type distribution tends to be equal, following the black diagonal distribution, with an average NCR $\gtrsim 0.4$ for every distribution. Type II, IIb, and Ia SN distributions begin to separate from the diagonal, and type Ibc, Ib, Ic, and IIn distributions begin to separate at bluer wavelengths. Also, type Ib SNe distribution gets closer to the diagonal as we move to bluer bands. The order Ia-II-IIb-IIn-Ibc remains mostly equal in all filters; however, IIn and IIb types present higher and lower NCRs in red bands, respectively.

We conducted Kolmogorov–Smirnov (KS) tests to compare two samples and assess the probability of both being drawn from the same probability distribution. By setting a significance level of 2$\sigma$, if the p value computed by the KS test falls below the threshold of 0.05, it means the rejection of the null hypothesis. In other words, it indicates they originated from different populations. Table \ref{tab:pvalues} shows the p value for every pair of NCR distributions and the diagonal hypothetical distribution that accurately traces the respective flux ([1:1]), and also for every filter. The table summarizes what is shown in Figure \ref{fig:ncr}. For the 12 J-PLUS filters, all distributions for filters redder than $J0410$ show p values > 0.05 (with the exception of the combination of II vs. Ibc types for $J0515$ and Ia vs. Ibc for $J0410$). For $uJAVA$, $J0378,$ and $J0395$, Ia and II-IIb-IIn shows p-value > 0.05. Only combinations of CC-types with Ia, and combinations of II with Ib, Ic and Ibc presents p-values < 0.05, indicating that types Ia, II, IIb, and IIn come from different possible progenitors than type Ib, Ic, and Ibc SNe
 that are located in redder environments in comparison with the rest of the types. 

The distributions of the five continuum-subtracted narrow-band filters are presented in Figure \ref{fig:ncr2}. We note again that SNe with $z > 0.014$ are not included in the \ion{N}{ii} distributions as explained in Section 3.5.
The most significant characteristic is the high number of zero NCR values in all cases. As we discuss later (Section 6), we associated part of the zero values found with the small aperture of the JAST80 telescope and the exposure times of the images used, which is not enough to collect enough photons to provide a high signal-to-noise ratio in these continuum-subtracted images for all galaxies. Moreover, H$\alpha$ is associated with stars of $M>15-20M_{\odot}$, so many SN progenitors are not expected to be associated with this emission, contributing to more zero NCRs in the distributions.
The removal of the [\ion{N}{II}] line has no appreciable effect on the NCR distributions, with the plots of H$\alpha+$[\ion{N}{II}], H$\alpha$ and [\ion{N}{II}] being similar.
In any case, we observe a similar scenario between the broad- and narrow-band filters, where the type Ia distribution is clearly the one with the lowest NCRs in all cases, followed by type II.
The [\ion{O}{II}] emission also traces the star formation rate \citep{1998ApJ...498..541K}, similarly to H$\alpha$.
%
With the change of signs in Ca II triplet fluxes, a higher NCR after subtracting the continuum suggests deeper absorption, which is analogous to high NCR values in emission-line filters and indicates strong emission (see Section 3.6). 
The order of the distributions appears to remain consistent, with Ia, II, and IIb displaying lower NCR values, distinguishing them from the other SN types. However, the large amount of zero values makes the analysis difficult in these cases.

\subsection{SN environment parameter distributions}

The local stellar population parameters $A_V$, $t_*$, SFR, and sSFR for all 418 SN environments are reported in Table \ref{tab:SSPresults}. Additionally, their distributions are illustrated in Figure \ref{fast++} and segregated according to the same seven SN groups.
These have been constructed similarly to in the previous section, that is, by performing 30000 different distributions where in each realization the best output values from the best FAST++ fits are randomly varied within their 1$\sigma$ uncertainty, and the median and the 1$\sigma$ variation of all realizations is taken as the final distribution.

We recover previous trends already found in the literature. 
The average local $A_V$ is larger for SNe Ibc ($1.13\pm0.08$ mag) compared to SNe Ia ($0.63\pm0.09$ mag), with the SNII distribution being in the middle of the two ($0.85\pm0.07$ mag), as previously reported by \cite{2017MNRAS.468..628G}. 
This trend is driven by the SNIb and SNIc distributions, which are clearly shifted to higher values compared to all others.
Similarly, all CC SNe distributions show younger average ages ($38\pm10$ Myr for SNe II and $43\pm13$ Myr for SNe Ibc) compared to those of SNe Ia environments ($222\pm73$ Myr),
and CC SNe environments have both higher SFR ($0.08\pm0.06$ and $0.16\pm0.04$ M$_\odot$yr$^{-1}$ for SNe Ibc and SNe II, respectively) and sSFR ($1.88\pm1.40$ and $4.35\pm1.29$ 10$^{-8}$ yr$^{-1}$ for SNe Ibc and SNe II, respectively) than SNe Ia ($0.01\pm0.01$ M$_\odot$yr$^{-1}$ and $0.19\pm0.16$ 10$^{-8}$ yr$^{-1}$) that occur in more passive local environments \citep{2018ApJ...855..107G}.

\begin{table*}[]
\resizebox{\textwidth}{!}{%
\begin{tabular}{cccccccccccccccccc}
\hline
Combination &
  uJAVA &
  J0378 &
  J0395 &
  J0410 &
  J0430 &
  gSDSS &
  J0515 &
  rSDSS &
  J0660 &
  iSDSS &
  J0861 &
  zSDSS &
  H$\alpha$+[\ion{N}{II}] &
  [\ion{O}{II}] &
  Ca II triplet &
  H$\alpha$ &
  [\ion{N}{II}] \\ \hline
II vs Ia &
  0.213 &
  0.740 &
  0.709 &
  0.757 &
  0.646 &
  0.896 &
  0.999 &
  0.442 &
  0.408 &
  0.997 &
  0.974 &
  0.973 &
  \multicolumn{1}{c|}{0.076} &
  \multicolumn{1}{c|}{0.013} &
  0.990 &
  0.399 &
  0.527 \\ \cline{2-4} \cline{15-15} \cline{17-18} 
\multicolumn{1}{c|}{II vs Ib} &
  \multicolumn{1}{c|}{0.002} &
  \multicolumn{1}{c|}{0.096} &
  \multicolumn{1}{c|}{0.023} &
  0.638 &
  0.392 &
  0.386 &
  0.083 &
  0.279 &
  0.169 &
  0.496 &
  0.632 &
  0.925 &
  0.095 &
  0.347 &
  \multicolumn{1}{c|}{0.158} &
  \multicolumn{1}{c|}{0.037} &
  \multicolumn{1}{c|}{0.017} \\ \cline{2-4} \cline{14-14} \cline{16-18} 
\multicolumn{1}{c|}{II vs Ic} &
  \multicolumn{1}{c|}{0.021} &
  0.164 &
  0.184 &
  0.438 &
  0.251 &
  0.810 &
  0.318 &
  0.777 &
  0.298 &
  0.421 &
  0.367 &
  \multicolumn{1}{c|}{0.391} &
  \multicolumn{1}{c|}{0.017} &
  \multicolumn{1}{c|}{0.461} &
  \multicolumn{1}{c|}{0.001} &
  \multicolumn{1}{c|}{0.001} &
  \multicolumn{1}{c|}{0.002} \\ \cline{2-4} \cline{8-8} \cline{14-14} \cline{16-18} 
\multicolumn{1}{c|}{II vs Ibc} &
  \multicolumn{1}{c|}{0.001} &
  \multicolumn{1}{c|}{0.037} &
  \multicolumn{1}{c|}{0.019} &
  0.194 &
  0.073 &
  \multicolumn{1}{c|}{0.289} &
  \multicolumn{1}{c|}{0.023} &
  0.229 &
  0.117 &
  0.399 &
  0.380 &
  \multicolumn{1}{c|}{0.495} &
  \multicolumn{1}{c|}{0.000} &
  \multicolumn{1}{c|}{0.134} &
  \multicolumn{1}{c|}{0.003} &
  \multicolumn{1}{c|}{0.000} &
  \multicolumn{1}{c|}{0.000} \\ \cline{2-4} \cline{8-8} \cline{14-14} \cline{16-18} 
II vs IIn &
  0.410 &
  0.321 &
  0.773 &
  0.695 &
  0.819 &
  0.343 &
  0.487 &
  0.211 &
  0.750 &
  0.814 &
  0.446 &
  0.434 &
  0.193 &
  0.364 &
  0.109 &
  \multicolumn{1}{c|}{0.069} &
  \multicolumn{1}{c|}{0.026} \\ \cline{18-18} 
II vs IIb &
  0.998 &
  0.918 &
  0.817 &
  0.899 &
  0.995 &
  0.985 &
  0.751 &
  0.695 &
  0.889 &
  0.805 &
  0.971 &
  0.987 &
  0.421 &
  0.848 &
  0.903 &
  0.859 &
  0.543 \\ \cline{2-4} \cline{14-15} \cline{17-18} 
\multicolumn{1}{c|}{Ia vs Ib} &
  \multicolumn{1}{c|}{0.000} &
  \multicolumn{1}{c|}{0.017} &
  \multicolumn{1}{c|}{0.021} &
  0.334 &
  0.117 &
  0.270 &
  0.135 &
  0.224 &
  0.173 &
  0.614 &
  0.777 &
  \multicolumn{1}{c|}{0.977} &
  \multicolumn{1}{c|}{0.000} &
  \multicolumn{1}{c|}{0.009} &
  \multicolumn{1}{c|}{0.256} &
  \multicolumn{1}{c|}{0.001} &
  \multicolumn{1}{c|}{0.002} \\ \cline{2-4} \cline{14-18} 
\multicolumn{1}{c|}{Ia vs Ic} &
  \multicolumn{1}{c|}{0.000} &
  \multicolumn{1}{c|}{0.027} &
  \multicolumn{1}{c|}{0.046} &
  0.104 &
  0.165 &
  0.770 &
  0.525 &
  0.669 &
  0.439 &
  0.704 &
  0.588 &
  \multicolumn{1}{c|}{0.740} &
  \multicolumn{1}{c|}{0.000} &
  \multicolumn{1}{c|}{0.003} &
  \multicolumn{1}{c|}{0.014} &
  \multicolumn{1}{c|}{0.000} &
  \multicolumn{1}{c|}{0.000} \\ \cline{2-5} \cline{14-18} 
\multicolumn{1}{c|}{Ia vs Ibc} &
  \multicolumn{1}{c|}{0.000} &
  \multicolumn{1}{c|}{0.003} &
  \multicolumn{1}{c|}{0.006} &
  \multicolumn{1}{c|}{0.046} &
  0.074 &
  0.217 &
  0.092 &
  0.342 &
  0.217 &
  0.506 &
  0.697 &
  \multicolumn{1}{c|}{0.959} &
  \multicolumn{1}{c|}{0.000} &
  \multicolumn{1}{c|}{0.000} &
  \multicolumn{1}{c|}{0.026} &
  \multicolumn{1}{c|}{0.000} &
  \multicolumn{1}{c|}{0.000} \\ \cline{2-5} \cline{14-18} 
Ia vs IIn &
  0.086 &
  0.184 &
  0.362 &
  0.449 &
  0.926 &
  0.155 &
  0.595 &
  0.271 &
  0.774 &
  0.832 &
  0.557 &
  \multicolumn{1}{c|}{0.557} &
  \multicolumn{1}{c|}{0.006} &
  0.076 &
  0.263 &
  0.014 &
  0.010 \\ \cline{14-15}
Ia vs IIb &
  0.627 &
  0.837 &
  0.700 &
  0.938 &
  0.999 &
  0.837 &
  0.893 &
  0.700 &
  0.999 &
  0.771 &
  0.996 &
  \multicolumn{1}{c|}{0.999} &
  \multicolumn{1}{c|}{0.018} &
  \multicolumn{1}{c|}{0.043} &
  0.965 &
  0.310 &
  0.121 \\ \cline{14-15}
Ib vs Ic &
  0.988 &
  0.999 &
  0.926 &
  0.999 &
  1.000 &
  0.879 &
  0.701 &
  0.890 &
  0.999 &
  0.994 &
  0.991 &
  0.984 &
  0.990 &
  1.000 &
  0.409 &
  0.380 &
  0.895 \\
Ib vs Ibc &
  0.967 &
  1.000 &
  0.975 &
  1.000 &
  0.999 &
  0.985 &
  0.924 &
  0.998 &
  0.999 &
  1.000 &
  1.000 &
  0.999 &
  0.989 &
  0.997 &
  0.998 &
  0.978 &
  0.995 \\
Ib vs IIn &
  0.806 &
  0.999 &
  0.889 &
  0.999 &
  0.405 &
  0.851 &
  0.987 &
  0.494 &
  0.999 &
  0.947 &
  0.967 &
  0.687 &
  0.903 &
  0.707 &
  0.947 &
  0.754 &
  0.754 \\
Ib vs IIb &
  0.131 &
  0.352 &
  0.862 &
  0.987 &
  0.553 &
  0.528 &
  0.221 &
  0.295 &
  0.271 &
  0.624 &
  0.760 &
  0.972 &
  0.964 &
  0.964 &
  0.770 &
  0.630 &
  0.630 \\
Ic vs Ibc &
  1.000 &
  1.000 &
  0.999 &
  1.000 &
  1.000 &
  0.999 &
  0.999 &
  0.991 &
  1.000 &
  0.999 &
  0.999 &
  0.999 &
  0.997 &
  0.999 &
  0.744 &
  0.827 &
  0.995 \\
Ic vs IIn &
  0.934 &
  0.997 &
  0.966 &
  0.999 &
  0.469 &
  0.731 &
  0.975 &
  0.789 &
  0.999 &
  0.995 &
  0.951 &
  0.951 &
  0.857 &
  0.707 &
  0.984 &
  0.681 &
  0.883 \\
Ic vs IIb &
  0.286 &
  0.349 &
  0.897 &
  0.835 &
  0.620 &
  0.793 &
  0.865 &
  0.512 &
  0.438 &
  0.280 &
  0.488 &
  0.512 &
  0.798 &
  0.964 &
  0.118 &
  0.179 &
  0.335 \\
Ibc vs IIn &
  0.965 &
  0.995 &
  0.927 &
  1.000 &
  0.466 &
  0.853 &
  0.999 &
  0.579 &
  0.999 &
  0.973 &
  0.903 &
  0.772 &
  0.569 &
  0.652 &
  0.999 &
  0.840 &
  0.813 \\
Ibc vs IIb &
  0.264 &
  0.393 &
  0.927 &
  0.957 &
  0.732 &
  0.673 &
  0.613 &
  0.393 &
  0.346 &
  0.443 &
  0.673 &
  0.788 &
  0.810 &
  0.880 &
  0.372 &
  0.314 &
  0.487 \\
IIn vs IIb &
  0.890 &
  0.688 &
  0.996 &
  0.996 &
  0.996 &
  0.551 &
  0.749 &
  0.284 &
  0.565 &
  0.714 &
  0.376 &
  0.578 &
  0.549 &
  0.873 &
  0.498 &
  0.553 &
  0.553 \\ \cline{2-18} 
\multicolumn{1}{c|}{II vs [1:1]} &
  \multicolumn{1}{c|}{0} &
  \multicolumn{1}{c|}{0} &
  \multicolumn{1}{c|}{0} &
  \multicolumn{1}{c|}{0} &
  \multicolumn{1}{c|}{0} &
  \multicolumn{1}{c|}{0.00038} &
  \multicolumn{1}{c|}{0} &
  \multicolumn{1}{c|}{0} &
  \multicolumn{1}{c|}{0} &
  \multicolumn{1}{c|}{0} &
  \multicolumn{1}{c|}{0} &
  \multicolumn{1}{c|}{0} &
  \multicolumn{1}{c|}{0} &
  \multicolumn{1}{c|}{0} &
  \multicolumn{1}{c|}{0} &
  \multicolumn{1}{c|}{0} &
  \multicolumn{1}{c|}{0} \\ \cline{2-18} 
\multicolumn{1}{c|}{Ia vs [1:1]} &
  \multicolumn{1}{c|}{0} &
  \multicolumn{1}{c|}{0} &
  \multicolumn{1}{c|}{0} &
  \multicolumn{1}{c|}{0} &
  \multicolumn{1}{c|}{0} &
  \multicolumn{1}{c|}{0.00125} &
  \multicolumn{1}{c|}{0} &
  \multicolumn{1}{c|}{0.00027} &
  \multicolumn{1}{c|}{0.00028} &
  \multicolumn{1}{c|}{0.00055} &
  \multicolumn{1}{c|}{0} &
  \multicolumn{1}{c|}{0} &
  \multicolumn{1}{c|}{0} &
  \multicolumn{1}{c|}{0} &
  \multicolumn{1}{c|}{0} &
  \multicolumn{1}{c|}{0} &
  \multicolumn{1}{c|}{0} \\ \cline{2-18} 
\multicolumn{1}{c|}{Ib vs [1:1]} &
  \multicolumn{1}{c|}{0.014} &
  \multicolumn{1}{c|}{0.001} &
  \multicolumn{1}{c|}{0.001} &
  \multicolumn{1}{c|}{0.003} &
  0.128 &
  0.357 &
  0.667 &
  0.107 &
  \multicolumn{1}{c|}{0.311} &
  \multicolumn{1}{c|}{0.035} &
  \multicolumn{1}{c|}{0.060} &
  \multicolumn{1}{c|}{0.019} &
  \multicolumn{1}{c|}{0.047} &
  \multicolumn{1}{c|}{0.056} &
  \multicolumn{1}{c|}{0.003} &
  \multicolumn{1}{c|}{0.100} &
  \multicolumn{1}{c|}{0.004} \\ \cline{2-5} \cline{11-11} \cline{13-16} \cline{18-18} 
\multicolumn{1}{c|}{Ic vs [1:1]} &
  \multicolumn{1}{c|}{0.002} &
  \multicolumn{1}{c|}{0.000} &
  \multicolumn{1}{c|}{0.001} &
  \multicolumn{1}{c|}{0.022} &
  0.071 &
  0.257 &
  0.069 &
  0.213 &
  0.339 &
  0.216 &
  0.144 &
  0.071 &
  \multicolumn{1}{c|}{0.295} &
  \multicolumn{1}{c|}{0.018} &
  0.278 &
  0.908 &
  0.193 \\ \cline{2-6} \cline{9-9} \cline{11-16} \cline{18-18} 
\multicolumn{1}{c|}{Ibc vs [1:1]} &
  \multicolumn{1}{c|}{0.000} &
  \multicolumn{1}{c|}{0.000} &
  \multicolumn{1}{c|}{0.000} &
  \multicolumn{1}{c|}{0.000} &
  \multicolumn{1}{c|}{0.000} &
  0.146 &
  \multicolumn{1}{c|}{0.016} &
  \multicolumn{1}{c|}{0.031} &
  \multicolumn{1}{c|}{0.036} &
  \multicolumn{1}{c|}{0.004} &
  \multicolumn{1}{c|}{0.002} &
  \multicolumn{1}{c|}{0.000} &
  \multicolumn{1}{c|}{0.011} &
  \multicolumn{1}{c|}{0.001} &
  \multicolumn{1}{c|}{0.000} &
  \multicolumn{1}{c|}{0.053} &
  \multicolumn{1}{c|}{0.000} \\ \cline{2-6} \cline{9-9} \cline{11-16} \cline{18-18} 
\multicolumn{1}{c|}{IIn vs [1:1]} &
  \multicolumn{1}{c|}{0.028} &
  \multicolumn{1}{c|}{0.021} &
  \multicolumn{1}{c|}{0.036} &
  \multicolumn{1}{c|}{0.146} &
  \multicolumn{1}{c|}{0.016} &
  0.855 &
  0.643 &
  0.952 &
  0.568 &
  0.366 &
  0.496 &
  0.496 &
  0.396 &
  0.473 &
  0.317 &
  0.711 &
  0.554 \\ \cline{2-6} \cline{9-9} \cline{11-18} 
\multicolumn{1}{c|}{IIb vs [1:1]} &
  \multicolumn{1}{c|}{0.000} &
  \multicolumn{1}{c|}{0.000} &
  \multicolumn{1}{c|}{0.001} &
  \multicolumn{1}{c|}{0.016} &
  \multicolumn{1}{c|}{0.005} &
  0.107 &
  \multicolumn{1}{c|}{0.017} &
  \multicolumn{1}{c|}{0.020} &
  \multicolumn{1}{c|}{0.068} &
  \multicolumn{1}{c|}{0.015} &
  \multicolumn{1}{c|}{0.023} &
  \multicolumn{1}{c|}{0.023} &
  \multicolumn{1}{c|}{0.037} &
  \multicolumn{1}{c|}{0.017} &
  \multicolumn{1}{c|}{0.000} &
  \multicolumn{1}{c|}{0.007} &
  \multicolumn{1}{c|}{0.001} \\ \hline
\end{tabular}%
}
\caption{p-values of the Kolmogorov-Smirnov test for every SN type combination and filter. We also included the p-values for the SN types and the diagonal hypothetical distribution that accurately traces the respective flux ([1:1]). Black font boxed values represent those distributions without an underlying causative relationship; i.e., we reject the null hypothesis that the two samples were drawn from the same 
probability distribution (p-value $<0.05$), and both NCR distributions come from different populations. The distribution used for the five continuum-subtracted filters have been rebuilt by proportionally removing the number of zero NCR values in each
distribution that correspond to the lowest observed fraction of zeros in an individual distribution (see Section 6).}
\label{tab:pvalues}
\end{table*}

\subsection{Radial distributions of the SNe}

To investigate potential correlations between the emission of a specific band and the distribution of SNe in galaxies, we calculated two parameters: the normalized galactocentric distance (NGCD) and the fraction of the r-band emission flux ($Fr$; \citealt{2009MNRAS.399..559A, 2010ApJ...717..342H}). Our approach involved fitting an elliptical 2D Gaussian model to the $rSDSS$ images of the galaxies to determine the shape and size of each galaxy. If the SN was located at the center of the ellipse, NGCD was assigned a value of 0; if the SN was positioned at the outer ellipse's edge, the NGCD equaled 1. We then fit a second ellipse, with the same parameters as the first one, but scaled to place the SN at the edge of this second ellipse. $Fr$ was subsequently computed as the ratio between the integrated fluxes in the rSDSS band of the second and first ellipses.

\begin{figure}
    \includegraphics[width=\columnwidth]{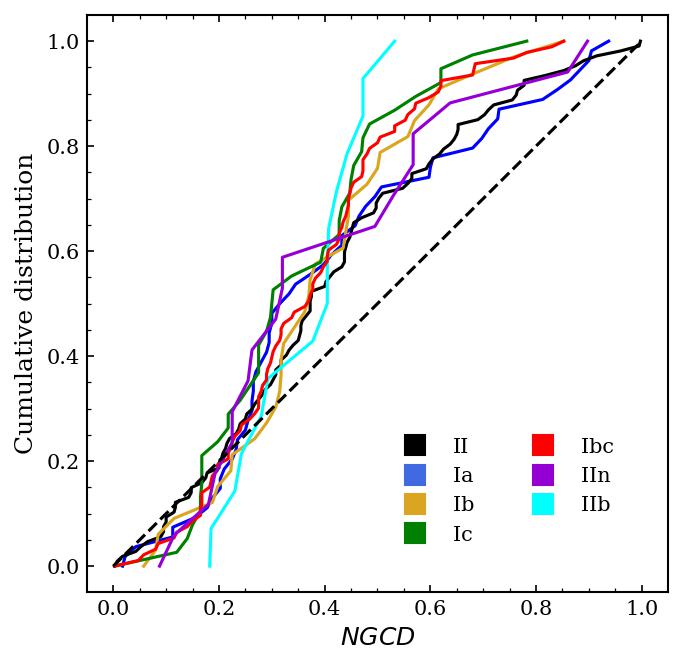}
    \includegraphics[width=\columnwidth]{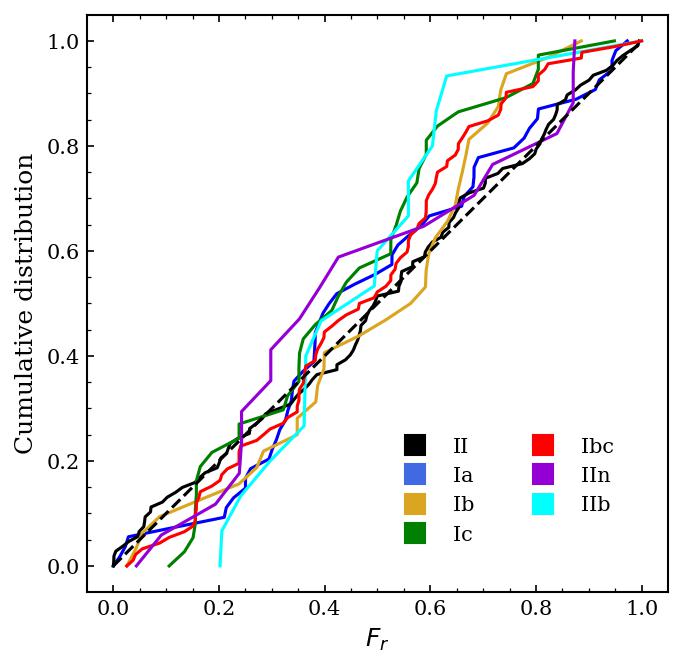}
    \caption{Cumulative distributions of normalized galactocentric distances (top) and F$_r$ (bottom) for every SN type.}
    \label{fig:GCDs}
\end{figure}

We find in Figure \ref{fig:GCDs} that all SN types exhibit nearly identical distributions with respect to each other, except for type IIb SNe, which appear to be concentrated in the inner regions of galaxies (within the range of 0.2 to 0.6 in NGCD). For the other SN types, there does not seem to be a preference for any specific location within the galaxies.

The $Fr$ are also similar among all SN types, and close to the diagonal, tracing the $rSDSS$ emission as the NCR distributions of Figure \ref{fig:ncr} do. Since all J-PLUS bands present the NCR distributions close to the diagonal, and there is no preference in the location of the SN, we only traced the general shape and size of the galaxies.

\section{Discussion}

\subsection{NCR results}

\begin{figure*}         
\centering
\includegraphics[width=0.32\textwidth]{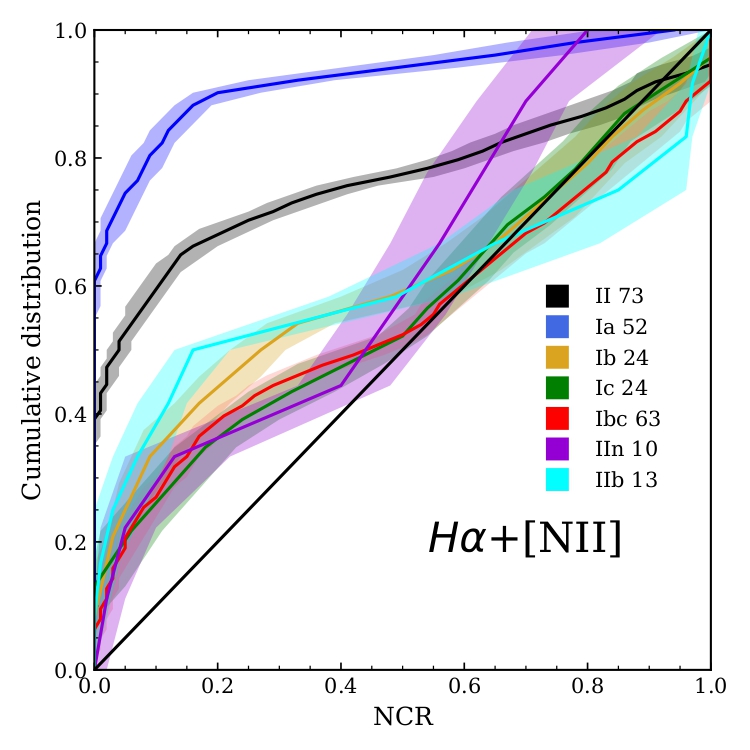}
\includegraphics[width=0.32\textwidth]{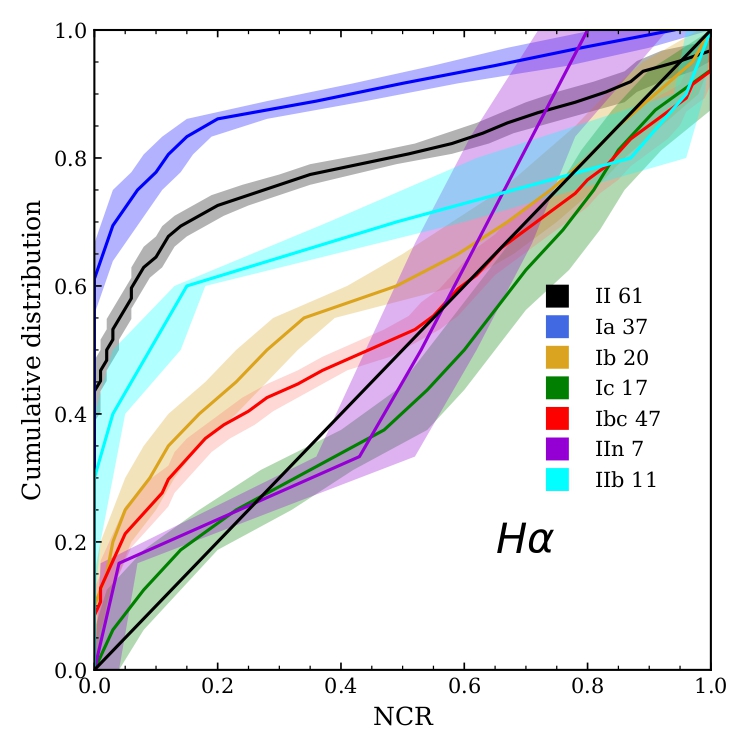}
\includegraphics[width=0.32\textwidth]{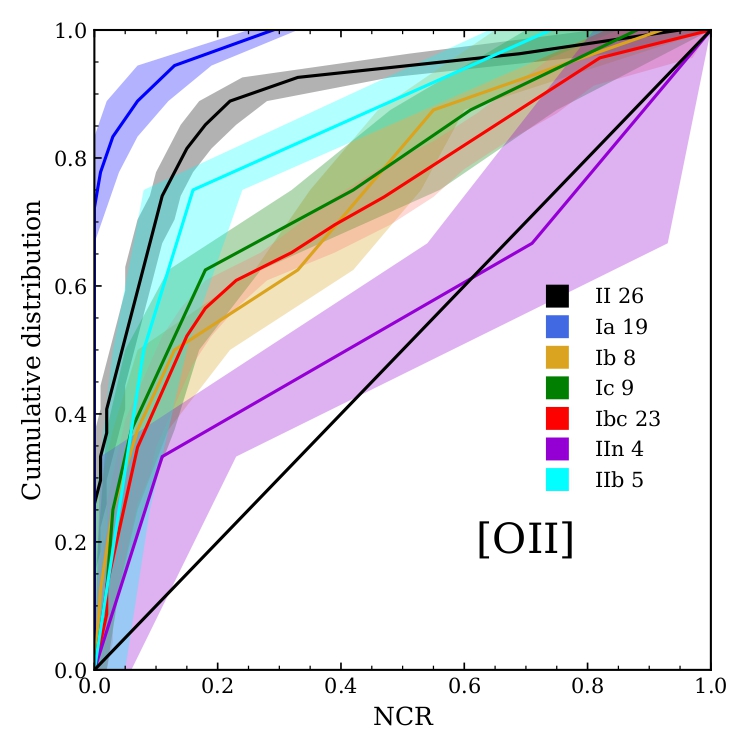}
\includegraphics[width=0.32\textwidth]{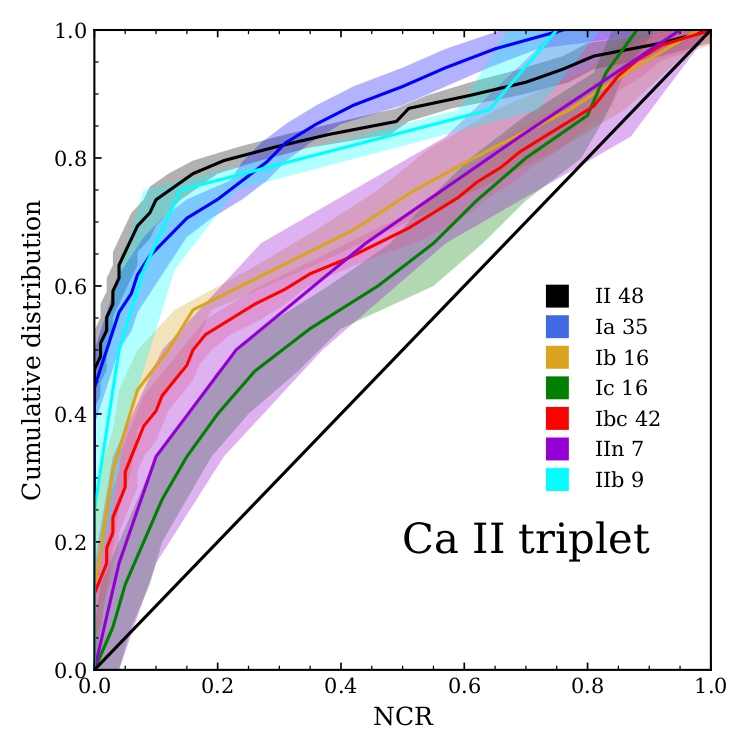}
\includegraphics[width=0.32\textwidth]{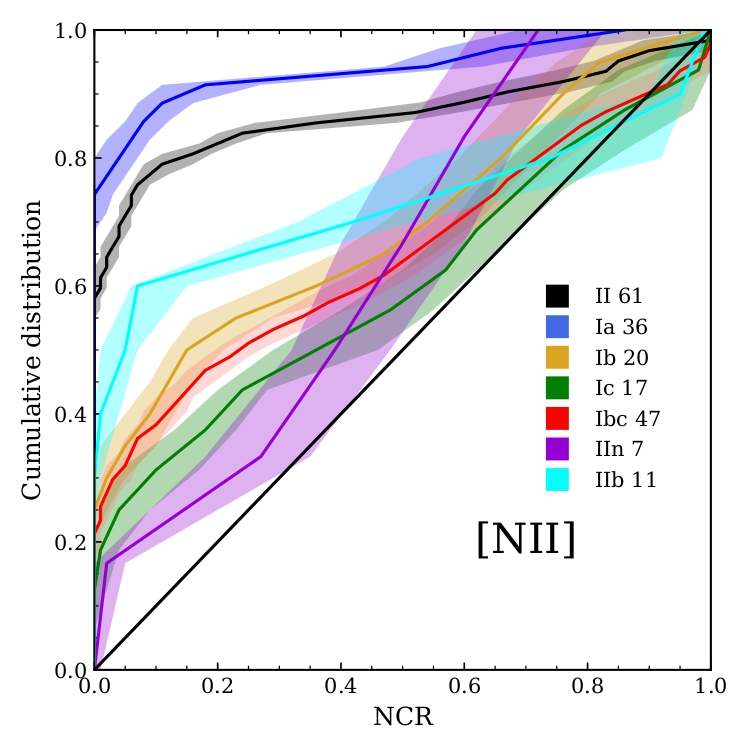}       
\caption{Same as Figure \ref{fig:ncr2}, but rebuilding the NCR distributions proportionally removing the number of zero NCR values in each distribution that correspond to the lowest observed fraction of zeros in an individual distribution.}
\label{no_zero_NCR}
\end{figure*}
\begin{figure*}         
\centering
\includegraphics[width=0.99\textwidth]{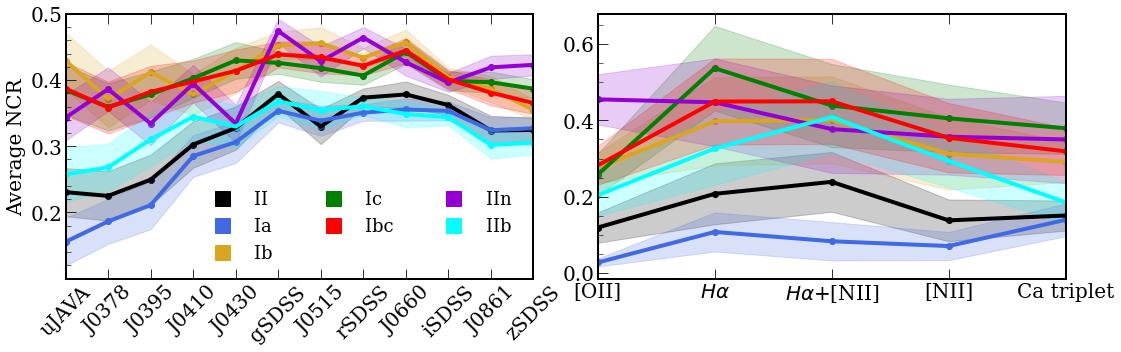}
\caption{Average NCR distributions as a function of the wavelength, for every SN type. The distributions are built averaging the NCR values for every J-PLUS photometric filter from the bluer to the redder filter (left panel). The right plot is built as the left one, but using the continuum-subtracted filters and proportionally removing the number of zero NCR values in each distribution that correspond to the lowest fraction of zeros in a distribution in each filter, as described in Section 6.1. }
\label{average_NCR}
\end{figure*}

Concerning the NCR distributions of the broad- and narrow-band filters and the average NCR values, it is worth noting that the broad filters exhibit slightly lower errors than the narrow-band filters, particularly for those where the continuum has been subtracted.
The errors for the $uJAVA$ filter are higher than the rest of the the broad-band filters.
This is what we expected since the broad filters have higher S/Ns, and the $uJAVA$ filter has the lower FWHM of all J-PLUS broad-band filters. 
We also observe approximately the same sequence between the main types in all panels in Figure \ref{fig:ncr}; this is type Ibc between II and Ia for bluer wavelengths, moving upward in the cumulative plot and downward for NCRs for redder wavelengths. This shift of the Ibc distribution is due to the Ib contribution, which correlates more with bluer bands, similarly to what was reported by \cite{2013MNRAS.436.3464K}. 
\cite{kelly} also found that the environments of SN Ib exists preferentially in environments with bluer surface brightness, especially in the ultraviolet band. In our case, we also found that the filter that best distinguishes the type Ib distribution is $uJAVA$. 

By looking at continuum-subtracted filters, the main characteristic is the large amount of zeros present in all distributions.
For instance, in the H$\alpha$ plot we have a noticeable amount of zero NCR values: $\sim 35\%$ for Ibc, $\sim 53\%$ for II, and $\sim 65\%$ for Ia. This is partly because many SNe do not follow the H$\alpha$ emission. However, this fraction of zeros is larger than in previous studies (e.g., \citealt{2012MNRAS.424.1372A} found $\sim 25\%$ for Ibc, $\sim 39\%$ for II, and $\sim 58\%$ for Ia).
Since these filters represent the residual flux once a pseudo-continuum (from broad-band filters) has been subtracted from a narrow-band filter, and since the diagonal line assumes a population that specifically traces the observed light, the distribution will include a fraction of zeros that depends on the image depth. Therefore, we have two possible explanations for this large amount of zeros; either most of the flux collected by the narrow-band filters is due to the underlying continuum, or the combination of a small telescope aperture (0.8m) and low exposure time was not enough to get significant contrast in these continuum-subtracted filters.

To further analyze this, we rebuilt the NCR distributions, this time proportionally removing the number of zero NCR values in each distribution that correspond to the lowest observed fraction of zeros in an individual distribution, following a procedure similar to \cite{2022MNRAS.513.3564R}.
This can be done as all our images are of a similar depth, since all the images were obtained using the same telescope with the same conditions, exposure time, and instrument configuration. This is equivalent to excluding the shaded region in all panels of Figure \ref{fig:ncr2}, because we assume we have reached the background level at that point, and tracing the diagonal from the intersection of the dotted horizontal line with the zero NCR to the upper right corner. This assumption is grounded in the fact that we successfully retrieved a certain level of signal, and anything below that threshold corresponds to the background we were not able to recover due to the combination of the telescope and exposure time used, as previously mentioned.
In Figure \ref{no_zero_NCR}, we show the resulting distributions. 
H$\alpha+$[\ion{N}{II}], H$\alpha,$ and [\ion{N}{II}] plots still show similar distributions; the main type that correlates the emission is Ic, with the highest NCR values on average (Figure \ref{average_NCR}). For the $J0660$ (Figure \ref{fig:ncr}) and H$\alpha$ (Figure \ref{no_zero_NCR}) panels for type Ia, II, and all Ibcs combined, we recovered the tendency of the distributions seen in previous works (e.g., \citealt{halphaem}), where type Ic/Ibc SNe is the closest distribution to the diagonal line, followed by type II and type Ia SNe. We interpreted this as a sequence in progenitor age and then inferred a sequence of progenitor mass, with the distributions closer to the diagonal being the SN type with higher mass progenitors \citep{2018A&A...613A..35K}.

Type IIn SN distribution also correlates with the H$\alpha$ emission, with lower NCR values than Ic ($\sim$0.05 on average), which is also consistent with the literature \citep{2022MNRAS.513.3564R}, with the same methodology), however, our nonzero IIn sample is reduced to only seven SNe.
All CC SNe types are strongly correlated to the [\ion{O}{II}] emission, that also traces SFR \citep{1998ApJ...498..541K} following the sequence of Ia-II-IIb-Ic-Ibc-Ib-IIn. If we exclude the bias in the IIn type, which is limited to only four SNe and not numerous enough to perform reliable statistics, we can observe that type Ib exhibits the most pronounced correlation with this emission, primarily because this type of SN is more closely associated with bluer bands, as previously mentioned. 
The same correlations between SN types are found for the H$\alpha+$[\ion{N}{II}], H$\alpha,$ and [\ion{N}{II}] filters. Besides, the Ia, IIb, and IIn types are correlated when the [\ion{N}{II}] is subtracted, the p-value being > 0.05 after the removal.

The Ca II triplet distribution makes a clear division between types II, IIb, and Ia and types Ib, Ic, and IIn (closer to the diagonal because of the change in the sign). 
These differences between populations are also evident when computing the p value of the KS test (Table \ref{tab:pvalues}): II versus Ic, II versus Ibc, Ia versus Ic, and Ia versus Ibc present p values < 0.05 for the Ca II triplet NCRs; II versus Ia, II versus IIb, and Ia versus IIb present p values > 0.05, as well as the rest of CC SNe.
The interpretation of the intensity of the Ca II triplet absorption is not trivial. Some works (e.g., \citealt{1989MNRAS.239..325D, 1998A&AS..130..513G}) have shown that the Ca II triplet equivalent width (EW) presents a degeneracy between stellar population age and metallicity. At high metallicity, the Ca II EW reaches its maximum value for stellar populations of $\sim$10 Myr, primarily attributed to the presence of red supergiant (RSG) stars. A secondary, lower peak emerges around 100 Myr, driven by stars in the asymptotic giant branch (AGB) phase. Furthermore, for ages over 1 Gyr, the Ca II EW becomes a reliable indicator of increasing metallicity.
Based on our results, and compared to SNII that explode in low Ca II NCR environments, one possible interpretation would be to attribute the higher values for Ib/Ic to young progenitors exploding at locations with metal-rich older populations. 
For SNIIn, although with low numbers, our results are in line with these progenitors being a combination of very young ($\sim$10 Myr peak) and older ($\sim$100 Myr peak) stars \citep{2018ApJ...855..107G}.

Finally, in Figure \ref{average_NCR} we show the average NCR values in all filters for the 7 SN types. On the left panel, we see that on average, through the 12 J-PLUS filters, SNe Ia present the lower NCR values (between $0.10\pm0.02$ and $0.33\pm0.02$), followed by II$<$IIb$<$IIn$<$Ibc, which are the highest NCR values for Ibc-Ib-Ic types, ranging between $0.36\pm0.01$ and $0.48\pm0.01$. The IIn oscillates among the Ia-II-IIb types and Ibc, obtaining values from $0.35\pm0.02$ up to $0.49\pm0.01$ in the $gSDSS$ an $rSDSS$ bands. Moreover, all NCRs of the Ia, II, and IIb types tend to increase as we move to redder wavelengths (up to 0.2 in NCR), while Ib, Ic, and Ibc tend to remain constant. 

For the five continuum-subtracted filters, shown in the right panel of Figure \ref{average_NCR}, Ic SNe tend to have higher NCR values on average in comparison with the corresponding J-PLUS bands, the average NCR being, for instance, $0.39\pm0.02$ in the$J0660$ filter and $0.54\pm0.05$ in H$\alpha$, and $\sim0.5$ for the rest, except the [\ion{O}{II}], whose average NCR falls to $0.26\pm0.02$. Moreover, the figure illustrates the low NCR values on average for Ia SNe in comparison with the rest of the SN types; these are especially low for [\ion{O}{II}]. The insignificant effect of the [\ion{N}{II}] removal from the $J0660$ filter for the rest of the SN types is also evident since the average NCR values for H$\alpha$ and H$\alpha+$[\ion{N}{II}] are similar in all distributions (except in type Ic, where the difference is $\sim$ 0.1). The IIn type is the next in terms of the highest NCRs. H$\alpha$ presents a value of $0.45\pm0.11$, which is consistent with the $0.521\pm0.038$ that \citet{2022MNRAS.513.3564R} found; however, we need to take into account that we only have seven type IIn SNe in our nonzero sample. The ordination of SN types between $J0378$ and [\ion{O}{II}] and between $J0861$ and Ca II triplet remains the same, but with systematically lower NCRs after subtracting the continuum.

\subsection{SED fitting}

\begin{figure}
                \includegraphics[width=\columnwidth]{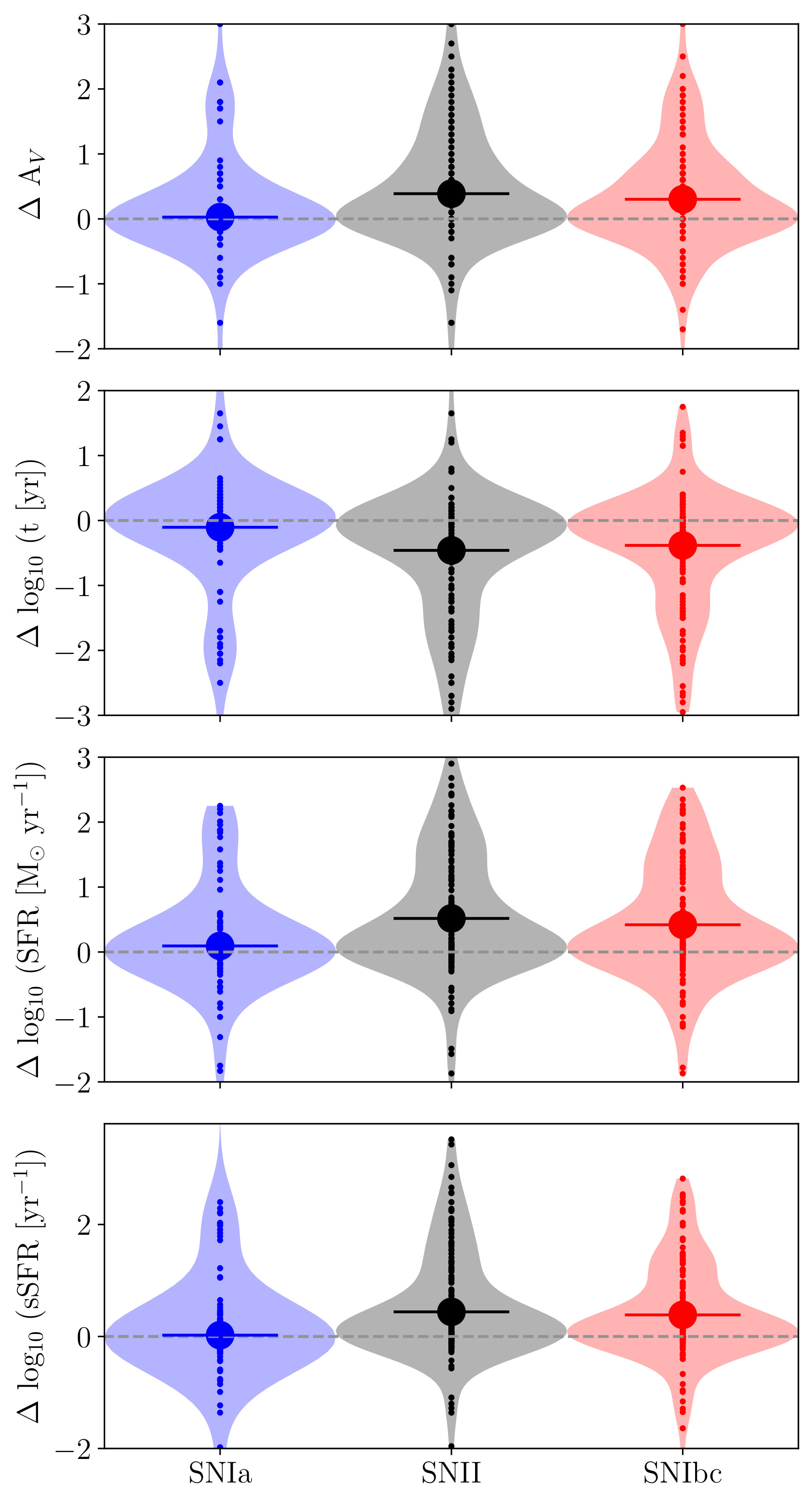}
        \caption{Difference between FAST++ SED fits once all 12 J-PLUS filters are used and those only using the five broad-band SDSS filters, for the four main environmental parameters. Individual dots represent the difference of each parameter for each individual SN environment, while the big dot represents the average of all differences. Violin plots represent the shape of the distributions.}
        \label{fig:diff}
\end{figure}

We recovered the same relations between the main SN types previously found in other works. 
Although the main focus of this work was to construct NCR distributions of the narrow-band filters, the fact that we recover these relations proves that the data is good enough to perform this kind of study.
However, to our knowledge, this is the first time that SED fitting is performed adding these seven narrow-band filters on top of the five main SDSS broad bands.

To see how much information these narrow-band filters carry that is not included in the broad bands, we repeated the FAST++ fits this time only using the five broad-band filters. 
In Figure \ref{fig:diff}, we show the differences between the twelve- and five-filter SED fits for the four main environmental parameters. Individual dots show the difference for each individual SN environment, the big dot represents the average of all differences, and the violin plots represent the shape of the distributions.
While all violins show the most populated value, similar to the median, compatible with zero difference, the average values show small shifts, which are more significant for the two CC SNe types with respect to Ia SNe. 
Given that the narrow filters are chosen so that they cover some of the the main emission lines in galaxy spectra, these shifts for CC SNe are expected because they occur next to HII regions that have intense emission lines. On the other hand, SNe Ia occur in all environments, most of them being non-star-forming without emission lines, so the narrow-band filters should not change the SED and the results of the fitting significantly.


\section{Conclusions}

We present a study of 418 environments of different supernova types using broad- and narrow-band images obtained by the J-PLUS collaboration with the JAST80 telescope. 
We measured the NCR value for each SN and constructed NCR distributions by SN type. 
Improvements over previous works include the extension of the original H$\alpha$ NCR analysis to other optical bands, the homogeneity of the data of those 418 environments using the same telescope, instrument configuration, spatial resolution, and image depth, and a better handling of the extinction correction and the stellar-continuum subtraction for the narrow-band filters. 
In addition, we constructed SEDs of all these nearby environments and performed SSP synthesis to estimate parameters such as star formation rate, stellar age, and extinction.

Regarding the NCR analysis on the J-PLUS filters, all broad-band distributions are quite similar, all of them being close to the diagonal. The fact that almost all broad-band distributions are close to the diagonal, along with the absence of a clear location preference for the SNe in the galaxies with respect to both NGCD and the fraction of $rSDSS$ flux (with the exception of the IIb type, which occur preferably in the inner part of the galaxies, between 0.2 and 0.5), suggests that we are primarily tracing the overall shape of the galaxy. The bluer bands ($uJAVA$, $J0378,$ and $J0395$) show the only difference with respect to the others; Ia, II, and IIb types present lower NCRs on average. When performing KS tests, the p values indicate that the underlying populations of Ia-II-IIb-IIn and Ibc-Ib-Ic are different, indicating that Ia-II-IIb progenitors trace the redder wavelengths better, also shown in the average NCR distributions.

The average NCR distributions for the five continuum-subtracted filters show a large fraction of zero values. For instance, in the H$\alpha$ emission we have  $\sim$35\% for Ibc, $\sim$53\% for II, and $\sim$65\% for Ia. This is partly because many SNe do not follow the H$\alpha$ emission. However, it is worth noting that there are more instances of zero values compared to previous studies \citep{2012MNRAS.424.1372A}. We attribute those extra zero values to the low signal-to-noise ratio in the narrow-band filters due to the combination of a small telescope aperture (0.8m) and low exposure time not being sufficient to obtain significant contrast in these continuum-subtracted filters. For this reason, the fraction of zero values is not directly comparable with other studies that use images with different depths.

After considering the lowest percent of zeros among the seven SN type distributions as our background, and proportionally removing the number of zero NCR values in each distribution that correspond to the lowest fraction of zeros in a distribution in each filter, we constructed new background-free NCR distributions, as performed by \cite{2022MNRAS.513.3564R}.
The SN Ic distribution correlates most strongly with H$\alpha$ emission. It is followed by IIn, Ibc, Ib, IIb, II, and Ia, showing that the progenitors of the Ic types are younger and more massive than type IIs, while Ia SNe progenitors are even older, recovering results shown by previous works. However, it is important to note that the distribution for type IIn is based on only seven SNe in this panel.
All core-collapse types strongly correlate with the [\ion{O}{II}] emission, which also traces SFR following the sequence of Ia-II-IIb-Ic-Ibc-Ib (excluding IIn since we have only 4 SNe). 
Besides this, the Ca II triplet makes a clear division between II-IIb-Ia and Ib-Ic-In progenitors, which we interpret as a difference in environmental stellar metallicity.
The KS test performed for the five continuum-subtracted filters also shows that II versus Ic, II versus Ibc, Ia versus Ic, and Ia versus Ibc come from different populations. The opposite occurs with II versus Ia, II versus IIb, and Ia versus IIb, as well as the
rest of CC SNe. In addition, after the [\ion{N}{II}] removal, the distinction between the underlying populations of the Ia, IIb, and IIn types, as revealed in the KS test, disappears, but the NCRs remain unchanged. We should note that the removal of [\ion{N}{II}] is oversimplified and may potentially introduce noise into this aspect of the analysis.

Regarding the SED analysis, we recovered the main relations previously reported among SN types for all parameters. Moreover, we found that including the J-PLUS narrow filters in the SED fitting has a more significant effect for the CC SNe environmental parameters, shifting all parameters due to the presence of strong emission lines in their typical star-forming regions.

In addition to the results that were known from the literature and that this paper confirms, the new findings presented can be summarized as follows:

\begin{itemize}
    \item[i)] NCR analysis for broad-band filters traces the overall shape of the galaxy.
    \item[ii)] IIb type SNe occur preferably in the inner part of the galaxies.
    \item[iii)] Ia, II, and IIb types occur preferably in redder environments.
    \item[iv)] CC SNe strongly correlate the [\ion{O}{II}] emission, tracing the SFR.
    \item[v)] High Ca II NCR environments can be interpreted as young progenitors exploding at locations with metal-rich older populations.
    \item[vi)] IIn progenitors are a combination of older and very young stars.
    \item[vii)] Including narrow filters in the SED analysis causes a shift in the four main environmental parameters for the CC SNe.
\end{itemize}

The JAST80 telescope with J-PLUS filters has allowed us to study SN host galaxies in the northern sky up to $z=0.0163$. The H$\alpha$ emission from galaxies at higher redshifts falls outside the width covered by the $J0660$ filter. In the near future, using the next generation of narrow-band wide-field surveys, such as J-PAS at the JST250 telescope \citep{2014arXiv1403.5237B} and the PAU survey at the William Herschel Telescope \citep{2019AJ....157..246P}, with filters covering a contiguous range at redder wavelengths, we will be able to extend the redshift range by using the filter covering the observed H$\alpha$ emission at $6563\times (1+z)$ \AA.

\begin{acknowledgements}

Based on observations made with the JAST80 telescope and T80Cam camera for the J-PLUS project at the Observatorio Astrof\'{\i}sico de Javalambre (OAJ), in Teruel, owned, managed, and operated by the Centro de Estudios de F\'{\i}sica del  Cosmos de Arag\'on (CEFCA). We acknowledge the OAJ Data Processing and Archiving Unit (UPAD) for reducing the OAJ data used in this work. Funding for OAJ, UPAD, and CEFCA has been provided by the Governments of Spain and Arag\'on through the Fondo de Inversiones de Teruel and their general budgets; the Aragonese Government through the Research Groups E96, E103, E16\_17R, E16\_20R and E16\_23R; the Spanish Ministry of Science and Innovation (MCIN/AEI/10.13039/501100011033 y FEDER, Una manera de hacer Europa) with grants PID2021-124918NB-C41, PID2021-124918NB-C42, PID2021-124918NA-C43, and PID2021-124918NB-C44; the Spanish Ministry of Science, Innovation and Universities (MCIU/AEI/FEDER, UE) with grant PGC2018-097585-B-C21; the Spanish Ministry of Economy and Competitiveness (MINECO) under AYA2015-66211-C2-1-P, AYA2015-66211-C2-2, AYA2012-30789, and ICTS-2009-14; and European FEDER funding (FCDD10-4E-867, FCDD13-4E-2685). The Brazilian agencies FINEP, FAPESP, and the National Observatory of Brazil have also contributed to this project.
The SNICE research group acknowledges financial support from the Spanish Ministerio de Ciencia e Innovaci\'on (MCIN) and the Agencia Estatal de Investigaci\'on (AEI) 10.13039/501100011033 under the PID2020-115253GA-I00 HOSTFLOWS project, from Centro Superior de Investigaciones Cient\'ificas (CSIC) under the PIE project 20215AT016, and the program Unidad de Excelencia Mar\'ia de Maeztu CEX2020-001058-M.
L.G. is also funded by the European Social Fund (ESF) "Investing in your future" under the 2019 Ram\'on y Cajal program RYC2019-027683-I.
RGB acknowledges financial support from the Severo Ochoa grant CEX2021-001131-S funded by MCIN/AEI/ 10.13039/501100011033.
Based on observations made with the JAST80 telescope/s at the Observatorio Astrofísico de Javalambre, in Teruel, owned, managed and operated by the Centro de Estudios de Física del Cosmos de Aragón, under proposals 1800146, 1900154, 1900165, 2000177, and 2000182.
We thank the OAJ Data Processing and Archiving Unit (UPAD) for reducing and calibrating the OAJ data used in this work. This work was funded by ANID, Millennium Science Initiative, ICN12\_009.

\end{acknowledgements}



\bibliographystyle{aa}
\bibliography{aa} 




\appendix

\section{Sample list and NCRs}

\small

\tablehead{\hline \textbf{NAME} & \textbf{TYPE} & \textbf{RA} & \textbf{DEC} & \textbf{z} \\ \hline}

\captionof{table}{Supernova sample. Name of the SN, type, right ascension, declination, and redshift.}

\begin{supertabular}{lllll}

SN2014ge              & Ib            & 181.2146    & 26.9963      & 0.0019     \\
SN2017igf             & Ia            & 175.7069    & 77.3702      & 0.0060     \\
SN2019clq             & Ib            & 114.1563    & 74.4465      & 0.0120     \\
SN2019tdf             & Ia            & 266.6207    & 30.7069      & 0.0155     \\
SN2019upq             & II            & 217.2838    & 27.4502      & 0.0143     \\
SN2019xdf             & II            & 141.9661    & 68.4119      & 0.0130     \\
SN2020aapr            & II            & 112.0179    & 66.3748      & 0.0150     \\
SN2020acac            & II            & 217.7526    & 25.4889      & 0.0151     \\
... \\ \hline

\label{tab:SNetable}\\
\multicolumn{5}{l}{\footnotesize \textbf{Note}: The full table is available at the CDS.}\\
\end{supertabular}

{\small

\tablehead{\hline \textbf{Name(Type)} & \textbf{uJAVA} & \textbf{gSDSS} & \textbf{rSDSS}&\textbf{iSDSS}&\textbf{zSDSS} \\ \hline}

\captionof{table}{NCR values for the broad-band filters.}

\begin{supertabular}{llllll}

ASASSN-14kg(II)  & 0.484 & 0.775 & 0.745 & 0.749 & 0.780 \\
ASASSN-15lf(II)  & 0.271 & 0.130 & 0.096 & 0.089 & 0.050 \\
ASASSN-15tw(II)  & 0.571 & 0.324 & 0.305 & 0.269 & 0.232 \\
ASASSN-16eu(II)  & 0.441 & 0.269 & 0.200 & 0.203 & 0.207 \\
ASASSN14az(IIb)  & 0.603 & 0.185 & 0.122 & 0.087 & 0.077 \\
ASASSN15bd(IIb)  & 0.819 & 0.661 & 0.576 & 0.578 & 0.569 \\
AT2020drl(II)    & 0.062 & 0.000 & 0.034 & 0.016 & 0.160 \\
... \\
\hline

\label{tab:NCRs_broad} \\
\multicolumn{5}{l}{\footnotesize \textbf{Note}: The full table is available at the CDS.}\\
\end{supertabular}

}

{\scriptsize
\tablehead{\hline \textbf{Name(Type)} & \textbf{J0378} & \textbf{J0395} & \textbf{J0410} & \textbf{J0430} & \textbf{J0515} & \textbf{J0660} & \textbf{J0861}  \\ \hline}

\captionof{table}{NCR values for the narrow-band filters.}

\begin{supertabular}{lllllllll}

ASASSN-14kg(II)  & 0.691 & 0.666 & 0.589 & 0.689 & 0.749 & 0.740 & 0.738 \\
ASASSN-15lf(II)  & 0.120 & 0.262 & 0.044 & 0.122 & 0.110 & 0.097 & 0.081 \\
ASASSN-15tw(II)  & 0.000 & 0.565 & 0.428 & 0.534 & 0.355 & 0.289 & 0.200 \\
ASASSN-16eu(II)  & 0.664 & 0.405 & 0.280 & 0.326 & 0.346 & 0.289 & 0.271 \\
ASASSN14az(IIb)  & 0.000 & 0.184 & 0.316 & 0.157 & 0.131 & 0.162 & 0.120 \\
ASASSN15bd(IIb)  & 0.821 & 0.773 & 0.732 & 0.736 & 0.707 & 0.623 & 0.551 \\
AT2020drl(II)    & 0.154 & 0.421 & 0.000 & 0.000 & 0.024 & 0.082 & 0.167 \\
... \\
\hline
\label{tab:NCRs_narrow}\\
\multicolumn{5}{l}{\footnotesize \textbf{Note}: The full table is available at the CDS.}\\
\end{supertabular}
}

{\tiny

\tablehead{\hline \textbf{Name(Type)} & \textbf{H$\alpha$+[NII]} & \textbf{[OII]} & \textbf{Ca II triplet} & \textbf{H$\alpha$} & \textbf{[NII]}  \\ \hline}

\captionof{table}{NCR values for the narrow-band filters after subtracting the corresponding underlying continuum.}

\begin{supertabular}{llllll}

ASASSN-14kg(II)  & 0.952 & 0.000 & 0.000 & 0.953 & 0.892 \\
ASASSN-15lf(II)  & 0.000 & 0.000 & 0.000 & 0.000 & 0.000 \\
ASASSN-15tw(II)  & 0.171 & 0.000 & 0.000 & 0.179 & 0.030 \\
ASASSN-16eu(II)  & 0.672 & 0.000 & 0.000 & 0.681 & 0.272 \\
ASASSN14az(IIb)  & 0.035 & 0.000 & 0.000 & 0.040 & 0.000 \\
ASASSN15bd(IIb)  & 0.753 & 0.584 & 0.000 & 0.759 & 0.659 \\
AT2020drl(II)    & 0.000 & 0.000 & 0.000 & 0.000 & 0.000 \\
... \\
\hline

\label{tab:NCRs_lines}\\
\multicolumn{5}{l}{\footnotesize \textbf{Note}: The full table is available at the CDS.}\\
\end{supertabular}

}

{\scriptsize
\tablehead{\hline \textbf{Name} &\textbf{Type} & \textbf{$\log <t_* (yr)>$} & \textbf{$A_V$ (mag)} & \textbf{$\log$ SFR} & \textbf{$\log$ M} \\ \hline}

\captionof{table}{Results of the SED fitting with all 12 filters.}
\begin{supertabular}{llllll}
SN2021yok &         II & 6.60$^{0.08}_{-0.13}$ & 2.10$^{0.03}_{-0.15}$ &  0.55$^{ 0.19}_{-0.23}$ &  6.85$^{ 0.07}_{-0.12}$ \\ 
SN2022prv &         II & 6.80$^{1.31}_{-0.17}$ & 0.80$^{0.41}_{-0.80}$ & -0.44$^{ 0.18}_{-1.29}$ &  6.06$^{ 0.05}_{-0.16}$ \\ 
SN1998ar &         II & 6.60$^{0.10}_{-0.11}$ & 1.40$^{0.03}_{-0.17}$ &  0.12$^{ 0.17}_{-0.27}$ &  6.42$^{ 0.07}_{-0.19}$ \\ 
SN1990L &         Ia & 7.65$^{0.05}_{-0.10}$ & 0.00$^{0.00}_{-0.00}$ & -3.31$^{ 0.07}_{-0.03}$ &  3.99$^{ 0.07}_{-0.04}$ \\ 
SN2019ltw &        IIb & 6.55$^{0.01}_{-0.30}$ & 1.40$^{0.05}_{-0.09}$ &  0.93$^{ 0.52}_{-0.01}$ &  7.18$^{ 0.24}_{-0.02}$ \\ 
SN2018cow &         Ic & 9.50$^{0.02}_{-2.47}$ & 0.00$^{1.43}_{-0.00}$ & -2.16$^{ 1.79}_{-0.02}$ &  6.87$^{ 0.02}_{-0.28}$ \\ 
SN2020acac &         II & 8.40$^{0.02}_{-1.78}$ & 0.00$^{1.50}_{-0.00}$ & -1.35$^{ 1.74}_{-0.03}$ &  6.72$^{ 0.05}_{-0.05}$ \\ 
... \\  
\hline
\label{tab:SSPresults}\\
\multicolumn{5}{l}{\footnotesize \textbf{Note}: The full table is available at the CDS.}\\
\end{supertabular}
}

\label{lastpage}
\end{document}